\documentclass[aps,prd,10pt,notitlepage,nofootinbib,superscriptaddress,showkeys,showpacs]{revtex4-1}

\usepackage{amsmath,amssymb,amsthm,latexsym} 

\usepackage{stmaryrd,wasysym,upgreek,mathrsfs,dsfont} 
\usepackage[english]{babel} 
\usepackage{graphicx,color} 
 
\usepackage{array}

\usepackage{hyperref}

\newtheorem{lemma}{Lemma}[section]

\newcommand{\al}{\underline{\alpha}}

\newcommand{\md}{{\rm d}}

\newcommand{\ind}{{\rm ind}_0}

\newtheorem{proposition}{Proposition}

\newtheorem{theorem}{Theorem}[section]

\newcommand{\beq}{\begin{equation}}
\newcommand{\eeq}{\end{equation}}
\newcommand{\be}{\begin{equation}}
\newcommand{\bee}{\begin{equation}}
\newcommand{\ee}{\end{equation}}
\newcommand{\bea}{\begin{eqnarray}}
\newcommand{\eea}{\end{eqnarray}}

\newcommand{\bal}{\begin{align}}
\newcommand{\eal}{\end{align}}

\newcommand{\cL}{{\cal{L}}}

\newcommand{\cG}{{\cal{G}}}
\newcommand{\cT}{{\cal{T}}}

\newcommand{\cF}{{\cal{F}}}

\newcommand{\cV}{{\mathcal V}}

\newcommand{\R}{\mathbb{R}} 
 
\newcommand{\Z}{\mathbb{Z}}

\newcommand{\s}{ \sigma }

\newcommand{\symc}{ \,{\rm sym}(1,2,3) } 

\newcommand{\bG}{{\partial\mathcal G}}

\newcommand{\pa}{{\partial}}

\newcommand{\tJ}{{\widetilde{J}}}

\newcommand{\col}{ {\rm{color}} }

\newcommand{\inter}{ {\rm{int\,}} }
\newcommand{\ext}{ {\rm{ext\,}} }

\newcommand{\bb}{{\mathbf b}}

\newcommand{\bmu}{\boldsymbol{\mu}}

\newcommand{\la}{\langle}
\newcommand{\ra}{\rangle}

\begin{document} 

 \title{A Renormalizable SYK-type Tensor Field Theory}  
\author{Joseph Ben Geloun}
\email{jobengeloun@lipn.univ-paris13.fr}
\affiliation{LIPN, UMR CNRS 7030, Institut Galil\'ee, Universit\'e Paris 13, Sorbonne Paris Cit\'e, 99, avenue Jean-Baptiste Cl\'ement, 93430 Villetaneuse, France, EU}
\affiliation{International Chair in Mathematical Physics and Applications, ICMPA-UNESCO Chair, 072Bp50, Cotonou, Benin}
\author{Vincent Rivasseau}
\email{vincent.rivasseau@gmail.com}
\affiliation{Laboratoire de Physique Th\'eorique, CNRS UMR 8627,
Universit\'e Paris XI,  F-91405 Orsay Cedex, France}

\begin{abstract} 
In this paper we introduce a simple field theoretic version of the Carrozza-Tanasa-Klebanov-Tarnopolsky (CTKT) ``uncolored" holographic tensor model. It gives a more familiar interpretation to the previously abstract modes of the SYK or CTKT models in terms of momenta. We choose for the tensor propagator the usual Fermionic propagator of condensed matter, with a spherical Fermi surface, but keep the CTKT interactions. Hence our field theory can also be considered as
an ordinary condensed matter model with a non-local and non-rotational invariant interaction. Using a multiscale analysis we prove that this field theory  is just renormalizable to all orders of perturbation theory in the ultraviolet regime.
\end{abstract} 

\maketitle 
\medskip


\medskip


\section{Introduction}
Holography (and in particular the AdS/CFT correspondence) provides 
an effective definition of quantum gravity systems dual to certain conformal field theories. However until recently the lack of simple solvable examples of this correspondence prevented to extract easily the gravitational content. 
A more serious shortcoming of AdS/CFT is that a second-quantized version of quantum
gravity should not be limited to a fixed space-time background, such as AdS. It should give a meaning to some kind of functional 
integral over space-times, presumably pondered by an action of the Einstein-Hilbert (EH) type. This seems up to now intractable in the continuum.

\medskip
Therefore, in parallel to string theory and AdS/CFT research, and largely independently from them, several formalisms 
have been developped in order to define a background-independent discretized
version of the quantum gravity functional integral. They go under various names such as dynamical and 
causal triangulations \cite{CDT}, group field theory \cite{GFT}, which is the second quantized version of loop quantum gravity \cite{Oriti:2014uga}, or random matrix and tensor models. The best success story in this direction is provided by 
random matrix models \cite{Di Francesco:1993nw}, for which the critical limit of 't Hooft topological expansion provides a universal 
random geometry \cite{browniansphere} now \emph{proven} equivalent to Liouville continuum gravity in dimension two 
\cite{Miller:2015qaa}.

\medskip
The Feynman graphs of random matrix models are dual to two dimensional triangulated surfaces.
Random tensor models of higher ranks were therefore introduced to perform a similar sum
but for higher dimensional triangulated geometries \cite{earlytensors}. They are indeed pondered by a discretized version 
of the EH action \cite{ambjorn}. But their development was impaired by the lack of analytic tools.

\medskip
Some years ago random tensors underwent a major upheaval. 
The theory was unlocked by the discovery of the tensor $1/N$ expansion \cite{tensor1/N}.
It provided the missing hierarchy for the Feynman graphs of tensor models. The leading order 
was identified as the now famous melonic family \cite{Bonzom:2011zz}. Surprisingly this melonic family
is \emph{simpler} than the planar family that leads 't Hooft expansion at rank two. But it is essential to add that the 
tensor $1/N$ expansion itself  (in its subleading orders) is \emph{much more complicated} than the
't Hooft expansion. At rank $d$  it organizes the
huge geometric category of piecewise linear quasi-manifolds of dimension $d$.
Several detailed reviews on this modern theory of random tensors are now available \cite{tensors}.
The corresponding revived approach  to quantum gravity forms the ``tensor track" \cite{tensortrack}.

\medskip
AdS/CFT correspondence and tensor models were until recently unrelated. This is no longer the case.
The Sachdev-Ye-Kitaev (SYK) model \cite{SYK}  provided two years ago a simple solvable
example of an ``almost" AdS$_2$/CFT$_1$ correspondence. It exhibits interesting properties such as maximal chaos 
\cite{Maldacena:2015waa}
and approximate conformal invariance, explicitly broken through a kind of bilocal BCS mechanism.
It is now clear that many details in the SYK model are not essential (Boson or Fermions, real or complex, particular rank etc...). The only feature which is not optional is the presence of \emph{at least one random tensor} which 
ensures that the large $N$ limit is governed by the melonic family. 

\medskip
The link between SYK and tensor models was made even tighter in the Gurau-Witten (GW) \cite{GW} and Carrozza-Tanasa-Klebanov-Tarnopolsky (CTKT) models \cite{Carrozza:2015adg,Klebanov:2016xxf}.
They open the new chapter of \emph{holographic tensor models}. 
All this research enjoys currently tremendous activity \cite{Klebanov:2017nlk,SYKtensors}. However there is one category
of random tensors still under the radar of the SYK and string community, namely \emph{tensor field theories} (TFTs) \cite{BGR,Geloun:2013saa,TFT,BenGeloun:2017vwn,BenGeloun:2017xbd}. 
TFT's distinguish themselves from tensor models by the presence of  a non-trivial propagator.
It allows to morph the $1/N$ limit into the physically more familiar picture of power counting, scales, and 
a renormalization group analysis, opening the possibility to search numerically for non-trivial fixed points \cite{RG}. 
Until now in SYK and holographic tensor models the modes are abstract and lack any spatial interpretation 
and the $N \to \infty$ limit is always performed at the beginning, keeping only the
leading $1/N$ terms. Remark that subleading effects in $1/N$ depend on the detail of the model chosen \cite{Bonzom:2017pqs}. In this way the $1/N$ limit cannot couple 
to the conformal limit. This seems to us somewhat unphysical. 

\medskip
In TFTs typical interactions still belong to the tensor
theory space \cite{Rivasseau:2014ima}
but the propagator (i.e. the Gaussian measure covariance)  is purposefully chosen to slightly break the tensor symmetry.
This is quite natural if we consider the tensorial symmetry as a kind of abstract generalization 
of  \emph{locality} in field theory \cite{Rivasseau:2012yp}. Propagators, as their name indicates, should \emph{break} locality.

\medskip
The main consequence of this slight breaking of the tensor symmetry is to allow for a separation 
of the tensor indices into (abstract, background-independent) ultraviolet and infrared degrees of freedom. 
Like in ordinary field theory most of the indices should have small covariances. They are identified with (abstract) ultraviolet
degrees of freedom. They should be integrated to compute the effective theory for the few indices which form the infrared, 
effective degrees of freedom (not the other way around!). This picture seems also related to the general AdS/CFT philosophy in 
which the renormalization group time, which flows 
between different conformal fixed points,
precisely provides the extra bulk dimension of AdS \cite{Ramallo:2013bua}.

\medskip 
At rank 2, TFT's reduce to non-commutative quantum field theory (NCQFT),
which is an effective regime of string theory \cite{Douglas:2001ba}. Mathematically it 
also corresponds to  Kontsevich-type matrix models instead of ordinary matrix models \cite{kontsevich}.
In the Grosse-Wulkenhaar version, it can be renormalized \cite{Grosse:2004yu} and the leading planar sector displays beautiful features such as 
asymptotic safety \cite{Disertori:2006nq} and integrability \cite{Grosse:2012uv}, together with a  completely unexpected restoration of Poincar\'e symmetry and of Osterwalder-Schrader positivity
\cite{GrosseRP}. 

\medskip
TFTs are the natural higher rank generalizations
of such NCQFTs.  When equipped with additional gauge projectors such TFT's coincide 
with tensor group field theory \cite{TGFT}, whose divergencies and radiative corrections 
require regularization, hence non-trivial propagators,  
as argued in \cite{Geloun:2011cy}.
An important unexpected property of TFTs is their generic asymptotic freedom, at least for 
quartic melonic interactions \cite{BenGeloun:2012pu,AF,Rivasseau:2015ova}.

\medskip

For all these reasons we introduce in this paper a first example of a tensor field theory of the
\emph{SYK-type}
\footnote{We could also call it a holographic tensor field theory, but we prefer to wait until
its holographic properties are better analyzed.}. 
The key is to choose an interesting propagator. Motivated by the condensed matter background of the SYK model,
we choose the usual propagator of Fermions in 4 dimensions with a spherical Fermi surface (jellium model
of non-relativistic many Fermions)\footnote{Therefore our model 
reminds of Horava-Lifschitz gravity or condensed matter physics, but beware that
the abstract ``space" of TFT's should not be necessarily identified
with ordinary coordinates on a semi-classical effective background.}, but we keep for interaction the two $O(N)^3$-invariant  
quartic tensor interactions of the Carrozza-Tanasa  \cite{Carrozza:2015adg} model. These interactions are the simplest
among all ``uncolored" \cite{Bonzom:2012hw} tensor interactions. Remark that the complete graph interaction 
has been also used in the context of the large $D$-limit of matrix models \cite{Ferrari:2017ryl} and recently
generalized to larger ranks in \cite{FRV}.

\medskip
In this paper we study the ultraviolet regime
of this model. Our main result is to prove a ``BPHZ-type" finiteness theorem 
at all orders through a multiscale analysis in the spirit of \cite{BGR,TGFT,Rivasseau:1991ub}.
We shall not discuss the non-perturbative stability here; see however \cite{constructiveTFT} for the 
constructive tensor field program, entirely devoted to this issue.

\medskip
The most interesting regime of the renormalization group in condensed matter physics is
governed by the low temperature excitations close to the Fermi surface. It is in this regime that we
expect to recover interesting holographic properties such as saturation of the maximal chaos bound \cite{Maldacena:2015waa}.
This requires a careful analysis in the style of \cite{condensed} which is left for a future study.

\medskip
Remark finally that our model is quite different from other types of tensor theories such as the Gross-Neveu tensor models studied in \cite{Prakash:2017hwq}
in which  the tensor invariance remains unbroken
by the propagator.

\section{The Model}
\label{sect:model}

\subsection{Fields}

Our goal is to extend into a tensor field theory the CTKT model \cite{Carrozza:2015adg,Klebanov:2016xxf}, using
the interactions of \cite{Carrozza:2015adg}, the time dependence {\it \`a la} SYK of \cite{Klebanov:2016xxf}, 
and a new propagator which mixes time with additional \emph{spatial} degrees of freedom.  Since we want to use
the Laplacian as our (non-relativistic) abstract spatial kinetic energy, and since it is a symmetric operator, we have first to double the number of fields.
So we consider \emph{a pair of Majorana tensor fields} which we write as 
$\{ \chi (t, \vec x, \sigma )\}$
where $\sigma$ is an abstract ``spin" index taking two values, 
1 or 2.\footnote{We could use the equivalent complex notation $\{ \psi (t, \vec x ), \bar \psi (t, \vec x )\}$ 
but this would take us further away from the initial SYK formalism.}
To stick for the moment as close as possible 
to the SYK and CTKT models we keep the interaction \emph{local in time}. But, and this is the defining feature 
of tensor field theory, our propagator is not local but has the ordinary form of a jellium condensed matter Fermionic propagator. 

The coordinates $\vec x$ replace the three $O(N)^3$-symmetric tensor indices. They
take value in a  Cartesian product $E^3$. In this paper we choose either $E= \R$, hence $\vec x =(x_1,x_2,x_3)\in  \R^3$, 
or a compactified version $E= U(1)$ and $\vec x =(\theta_1,\theta_2,\theta_3)\in U(1)^3$, the three dimensional torus. 
Remember, however, not to  identify this $\vec x$ variable with an ordinary direct space coordinate, 
as the CT interaction is neither rotation invariant nor local in 
terms of these variables. 

The time variable is taken on the thermic circle $[- \frac{\beta}{2}, \frac{\beta}{2}]$. 
Since $\beta= \frac{1}{kT}$, this thermal circle becomes large
at low temperature. 
We also introduce the dual momentum variables $(p_0,\vec p)$. $p_0$, often called $\omega $ in condensed matter and SYK literature, is a Euclidean Matsubara frequency, hence it takes values in a $\Z$ lattice of small mesh $ \frac{2 \pi  }{\beta}$; if $\chi$ is Fermionic we should take
anti-periodic conditions, which, since $p_0 = \frac{2 \pi  }{\beta} (n+ \frac{1}{2})$, provide a natural infrared cutoff. This will not important in the subsequent analysis where $p_0$ is taken large compared to the lattice spacing.

Similarly the momenta dual to the $\vec x$ variables will be denoted generically as $\vec p$. They take values in $ \R^{3}$ or $\Z^3$
depending upon whether we choose $E= \R$ or $E = U(1)$, but this is again quite irrelevant for our analysis which considers 
a regime of the theory at large $\vec p$. We introduce the notations 
$p^2= |\vec p\,|^2=\sum_{i=1}^3 p_i^2$, 
and $\int \md^3 p $ means either $ \int_{\R^3}\md p_1\md p_2 \md p_3$
in the non-compact case $E = \R$ or $ \sum_{(p_1, p_2, p_3 )\in \Z^3} $ in the compact case $E = U(1)$. The difference is not essential since in this paper we shall study the theory at large momenta only.

\subsection{The propagator}

Using the Matsubara formalism and the notations of \cite{DMR},
the propagator in Fourier space $\underline{\hat{C}}$ of a condensed matter Fermionic field 
living on space ${\mathbb R}^3$  at finite temperature $T$ is equal to:
\begin{equation}
\underline{\hat{C}} (p_0, \vec p) =     \frac{1}{ip_0-e(p)},
\quad \quad e(p)= \frac{p^2}{2m}-\mu \ ,
\label{prop}
\end{equation}
where the vector $\vec p$ in (\ref{prop}) is three-dimensional, 
and the parameters $m$ and $\mu$ correspond to the effective mass and to
the chemical potential (which fixes the Fermi energy).
To simplify we put for the moment $2m= \mu=1$, so that
$e(p)=  p^2-1$. 
The corresponding direct space propagator at temperature
 $T$ and position $(t, \vec x)$
(where $\vec x$ is the three dimensional spatial component) is 
\begin{equation}
\underline{C}(t, \vec x) =\frac{T}{(2\pi)^3}\; \sum_{p_0 } \; \int \md^3p\; e^{-ip_0 t +ip \cdot x}\;
\underline{\hat{C}}(p_0, \vec p) \ .
\label{tfprop}\end{equation}
It is antiperiodic in the variable
$t$ with antiperiod $\frac{1}{T}$. This means that
\begin{equation}
\underline{\hat{C}} (p_0, \vec p)= \frac{1}{2}\int_{-\frac{1}{T}}^{\frac{1}{T}} \md t
\int \md^3x \;e^{+ip_0 t -ip \cdot x}\; \underline{C}(t, \vec x)
\end{equation}
is not zero only  for   discrete
values (called the Matsubara frequencies) :
\begin{equation}
p_0 =   (2n+1) \pi T \ , \quad n \in {\mathbb Z} \ , 
\label{discretized}
\end{equation}
where we take $ \hbar =k =1$. Remark that only
odd frequencies appear, because of  antiperiodicity, 
hence $\vert p_0\vert  \geq \pi T$
so that the temperature acts like an effective infrared cutoff.

The notation $\sum_{k_0}$ in (\ref{tfprop})
means really the discrete sum over the integer
$n$ in (\ref{discretized})\footnote{When $T \to 0$,  $k_0$
becomes a continuous variable, the discrete sum becomes an
integral  $T\sum_{k_{0}}\rightarrow \frac{1}{2\pi }\int \md k_{0} $, 
and the corresponding propagator
 $C_{0}(k_0, \vec k)$ becomes singular
on the Fermi surface
defined by $k_0=0$ and $|k|=1$.}.
To simplify notations we write:
\begin{equation}
\int \md^4p \; \equiv \; T\sum_{p_0} \int \md^3p
\; , \quad 
\int \md^4x \; \equiv \; \frac 12
\int_{-1/T}^{1/T}\md t \int \md^3x \ . \label{convention}
\end{equation}

\be
\underline{\hat{C}} (p_0,\vec p):=  
 \frac{-ip_0 +\ e(p)  }{p^2_0 + e^2(p) } 
 = 
 \int_{0}^{\infty} \md\alpha \, (-ip_0 +\ e(p) )  e^{-\alpha  (p^2_0  + e^2(p))}
  \label{proparam1}.
  \ee

To study the ultraviolet regime of the theory we 
can consider only large values of $p_0$ and $e(p)$. In that regime we can write
\be
\underline{\hat{C}}(p_0,\vec p):=  
 \frac{-ip_0 +\ e(p)  }{p^2_0 + e^2(p) } (1- e^{- (p^2_0  + e^2(p))})
 = 
 \int_{0}^{1} \md\alpha \, (-ip_0 +\ e(p) )  e^{-\alpha  (p^2_0  + e^2(p))}
 \label{proparam2}.
\ee

We then adopt  then the following covariance for our free model with abstract spin
is defined by the matrix covariance rules
\bea
\label{propagChi}
&& 
\Big(\langle \chi_\s (p_0, \vec p) \chi_{\s'} (p_0', \vec p\,') \rangle \Big)_{\s\s'}= \Big(C_{\s\s'} (p_0, \vec p) \delta(p_0-p_0')\delta(\vec p,\vec p\,')\Big)_{\s\s'} \cr\cr
&&
= \left[\frac{ip_0}{p_0^2 + e^2( p^2) }  \begin{pmatrix} 1 &0 \cr 0 &1 \end{pmatrix} +   \frac{e( p^2 )}
{p_0^2 + e^2( p^2) }  \begin{pmatrix} 0 &-1 \cr 1 & 0 \end{pmatrix}\right]  \delta(p_0-p_0')\delta(\vec p,\vec p\,')\,, 
\eea
where $\chi_\s(p_0,\vec p) = \chi(p_0,\vec p,\s),$ and 
$\s, \s'$ are the spin indices, and the matrices refer to these indices. Remark that these rules are globally antisymmetric,
as they should be for Grassmann variables.

Denoting $d\mu_C (\chi) $ the corresponding Grassmann Gaussian measure \cite{condensed}
the free theory is defined with $J_\s$ a Fermionic tensor source field (also with a two-valued spin index) and $J \cdot \chi =\sum_\s  \int \md p_0\md^3p\, 
J_\s (p_0,\vec p) \chi_\s(p_0,\vec p)$. $C$ is the covariance of the Gaussian
measure, or free propagator and $\int d \mu_C$ is the Gaussian integral of
 covariance $C$. We are interested in computing the partition function
 $Z$\bea\label{partition}
Z(J) =
\int \md\mu_{C}(\chi) e^{ - S[\chi] + J \cdot \chi } \,, 
\eea and the generating function for cumulants of the theory
\be Z(J) =  W (J) = \log Z (J) \, .
\ee

\subsection{The tensor interaction}

We equip the free model with interactions inspired by  those of  
Carrozza-Tanasa\footnote{The first term of this interaction 
with coupling $\lambda_+$ is also the one used by F. Ferrari for the large $D$
limit of matrix models \cite{Ferrari:2017ryl}.} \cite{Carrozza:2015adg}. 
The tetraedric part of that interaction was used also in 
\cite{Klebanov:2016xxf}. 

\begin{figure}[t]
  \begin{center}
  \includegraphics[width=0.3\textwidth]{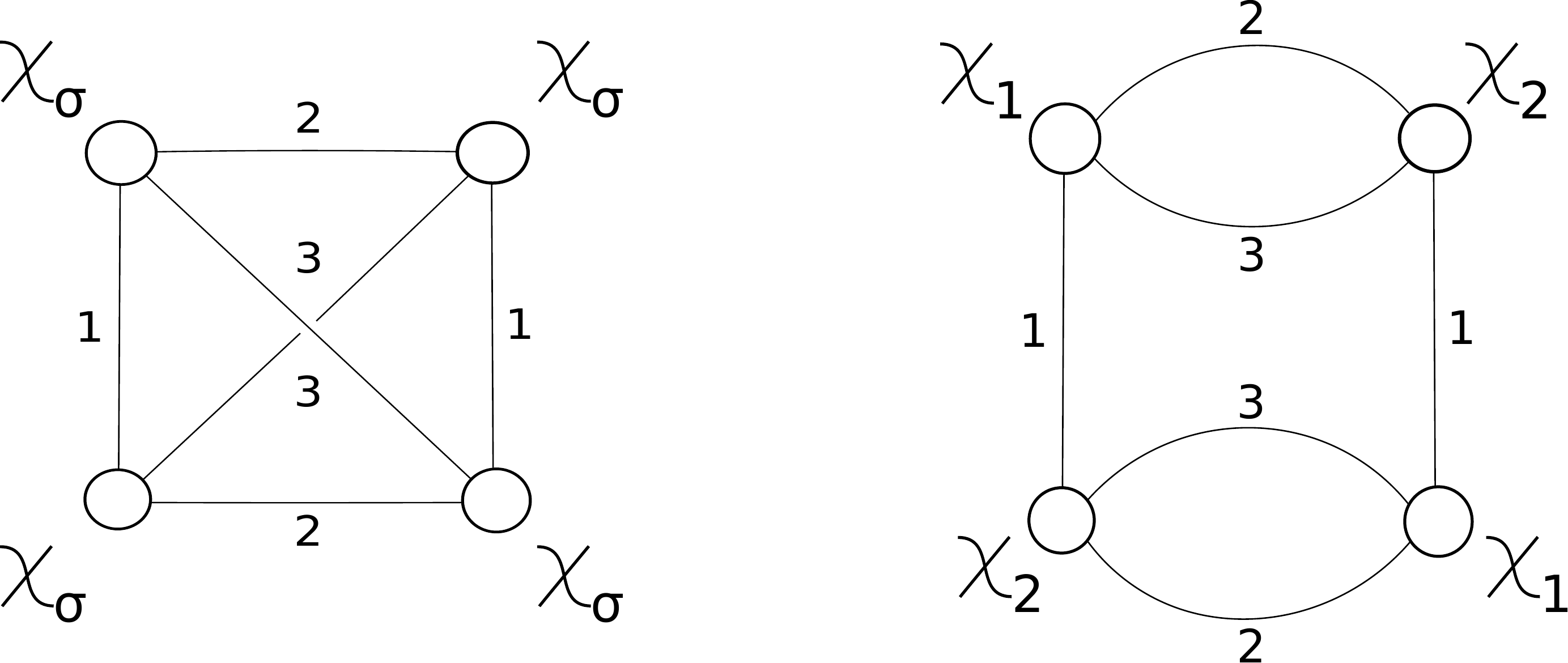}
   \end{center}
  \caption{$O(N)$ invariants as interactions: on the left, tetraedric invariant associated with 
  $\bb_+$; on the right, le spheric melonic invariant  $\bb_1$.}
  \label{oninvar}
\end{figure}

Consider the following interaction, 
\beq\label{sinter}
S_{\inter}(\chi)= \lambda_+ I_{\bb_+}(\chi)  + \lambda_m \sum_{c=1}^3I_{\bb_c}(\chi) + V_2 (\chi )
\,,
\eeq
where the coupling constants $\lambda_+$ and $\lambda_m$ ($m$ standing for ``melonic") are the bare coupling constants (which themselves decompose into renormalized constants plus counterterms) and where $I_{\bb_+}$ and $I_{\bb_c}$ are the quartic interaction terms
fully expanded in $(p_0,\vec p)$-space representation as
\bea
&& 
I_{\bb_+} = \sum_{\s=1,2} 
 \int  [\prod_{l=1}^4 \md p_{0;l}] \md^3p \md^3 p' \,
\chi_{\s} (p_{0;1}, p_1, p_2, p_3) \chi_{\s} (p_{0;2}, p_1, p'_2, p'_3) \chi_{\s} (p_{0;3}, p'_1, p_2, p'_3) 
\chi_{\s} (p_{0;4}, p'_1, p'_2, p_3)  \delta(\sum_{l=1}^4 p_{0;l}) \,,\crcr
&&
\sum_{c=1}^3I_{\bb_c}(\chi)  =  \int 
 [\prod_{l=1}^4 \md p_{0;l}] \md^3 p \md^3 p'\,
\chi_1 ( p_{0;1}, p_1, p_2, p_3) \chi_2 ( p_{0;2}, p_1', p_2, p_3) \chi_1 ( p_{0;3}, p'_1, p'_2, p'_3) \chi_2 ( p_{0;4}, p_1, p'_2, p'_3) 
\delta(\sum_{l=1}^4 p_{0;l}) \crcr
&& 
 \qquad \qquad\quad  + \symc    \,, \label{act:tens}
\eea
where the integrals are understood as \eqref{convention}, and \symc replaces the sum over colors.
The two types of interactions are associated with bubble diagrams $\bb_+$ and $\bb_c$
which represent orthogonal invariants as depicted in Figure \ref{oninvar}. 
Note that in this figure, only the melonic bubble $\bb_1$ is drawn and the other
bubbles with colors 2 and 3 can easily recovered.

Melonic interactions  \cite{Bonzom:2011zz,tensors} belong to the family of dominant terms at  large $N$,
and we expect that they will be dominant in the ultra-violet regime. 
Remark also that the above interactions are local in the $p_0$-space
but non local in the $\vec p$-space. 
Attached to the local variables, a delta function $ \delta(\sum_{l=1}^4 p_{0;l})$ 
at each vertex  manifests the conversation of momenta entering and
exiting from the vertex.  This is the usual standard of quantum field theory. 

To be more specific, the most natural rule at this stage seems to keep the tetraedric interaction in $\lambda_+$  \emph{diagonal in spin indices}, hence the sum over $\s$ of an
 interaction with four $\chi_\s$'s, as if we had two independent Majorana fields. 
However for the melonic interaction, since it is bipartite
we feel the most natural interaction is to mix the spins hence to choose two spins and two anti spins cyclically along the melonic cycle, see Figure \ref{oninvar} which shows 
the vertices associated with these interactions. 
These specific spin index choices for the interactions
could be modified if that leads to more interesting infra-red physics.

The remaining term in \eqref{sinter}, $V_{2}$ 
gathers the two-point function mass and wave function counterterms:
\bea\label{v2}
V_2(\chi) &=&  \Delta \mu(\lambda_m, \lambda_+) \sum_{\s<\s'}\int \md p_0 \md^3 p\;   \chi_{\s}  (p_0,\vec p) \chi_{\s'} (-p_0,\vec p)  
+\Delta_{p_0} (\lambda_m, \lambda_+)\sum_{\s} \int \md p_0 \md^3 p\; (ip_0)   \chi_{\s}  (p_0,\vec p) \chi_{\s} (-p_0,\vec p) \cr\cr
&+& \Delta_{ p^2} (\lambda_m, \lambda_+)
 \sum_{\s<\s'} \int \md p_0 \md^3 p\;    p^2 \chi_{\s}   (p_0,\vec p) \chi_{\s'}(-p_0,\vec p)  .
\eea
In this formula, as usual in perturbative renormalization, 
the mass counterterm $  \Delta \mu(\lambda_m, \lambda_+)$ and wave function counterterms 
$ \Delta_{p_0} (\lambda_m, \lambda_+)$ and $ \Delta_{ p^2} (\lambda_m, \lambda_+)$ are themselves
perturbative series in the coupling constants.

\begin{figure}[h]
  \begin{center}
  \includegraphics[width=0.4\textwidth]{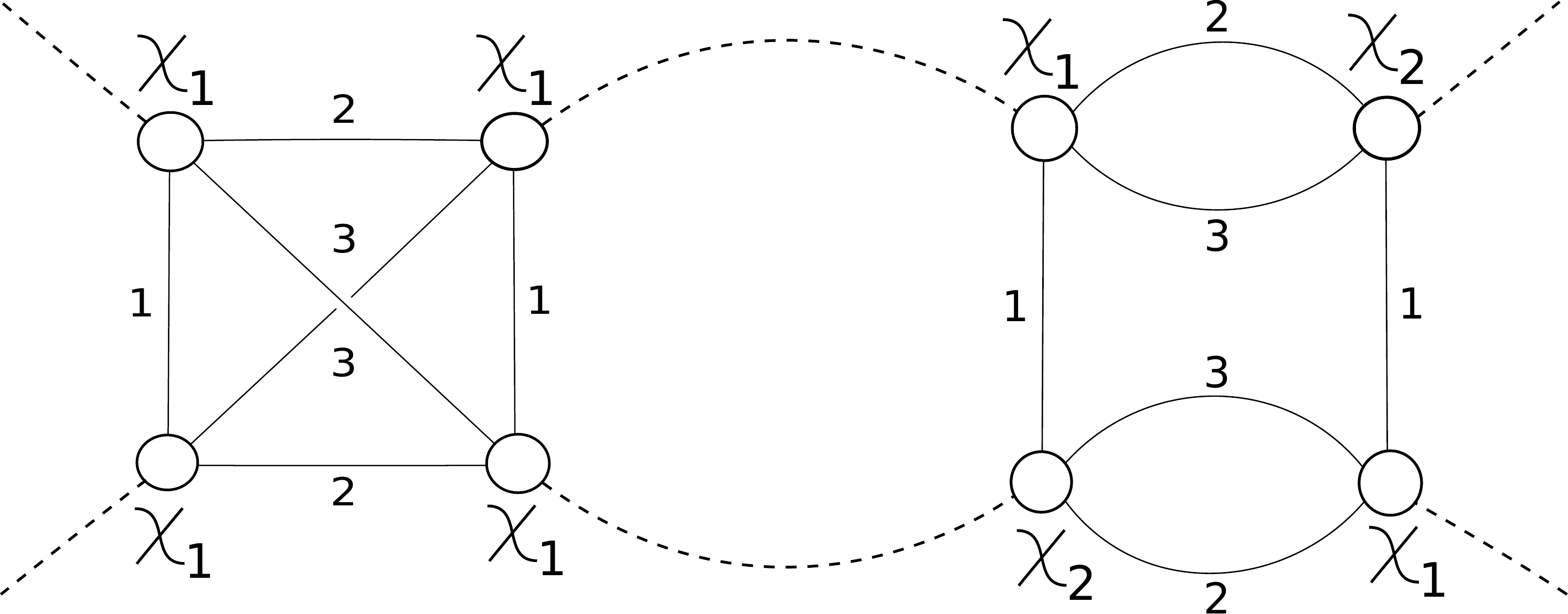}
   \end{center}
  \caption{A Feynman graph in the theory.}
  \label{onFG}
\end{figure}

A priori the counterterms could be power series in both couplings but we shall see below that only the melonic vertex is relevant
in the ultraviolet regime. We nevertheless also included the tetraedric vertex because we feel it is the one which could be responsible for SYK physics in the infrared regime. 
Finally, a Feynman graph in this theory is formed with the gluing of vertices 
$\bb_+$ and $\bb_c$ with propagator lines that we draw as dashed lines in order
to distinguish them from the internal structure of the vertices. See Figure \ref{onFG}. 
As one quickly understands, a Feynman graph in this setting  is a 4-regular edge (line) colored graph with half-lines. The propagator lines will be associated with the color 0.

\subsection{Amplitudes}

Expanding the theory in Feynman graphs, the amplitudes 
have to be arranged as Pfaffians of the antisymmetric matrix $C$ \cite{condensed} and have the general form 
\bea
&&
\la \prod_{a=1}^q\psi(p_{0;a},\vec p_a) \ra
 = \sum_{n=0}^\infty \frac{1}{n!} \int \md\mu_C(\psi) \Big[\prod_{a=1}^q\psi(p_{0;a},\vec p_a) \Big]\Big[-S_{\inter} (\psi)\Big]^n
 =  \sum_{\cG}   
 A_{\cG} \,. 
\eea
In the above expression, $\psi$ stands either for  $\chi_1$ or for $\chi_2$. 
As already emphasized, the spin index does not matter in the ultraviolet study, but can strongly affect 
the infra-red regime. 

Feynman amplitudes $A_{\cG}$  will be our focus. 
For the moment and for simplicity, we neglect the presence of mass and wave-function counterterms. 
They will be discussed in the following sections. Therefore we 
consider a connected amputated graph $\cG$ with
  vertex set 
 $\cV= \cV_+ \cup \cV_{m}$, 
 with cardinal $V= \vert \cV \vert $,
 where 
$\cV_+$ is the set of tetraedric vertices with pattern $\bb_+$
and $\cV_m$ is the set of melonic vertices $\bb_c$,
and with line set $\cL$, with cardinal  $L= \vert \cL \vert $.
Note that $\cL$ decomposes in two sorts of lines: $\cL_1$ associated
with the diagonal part of the covariance $C_{\s\s}$, and  $\cL_2$
associated with the off-diagonal entries. 
We denote $N_{\ext}$ the number of external fields also called external  legs.
Henceforth, the index $\s$ will be mostly omitted in the notations
but their presence  is however indicated by the two sets $\cL_i$, $i=1,2$. 

The bare amplitude of a Feynman graph $\cG$ is given by 
\bea\label{ampl0}
A_{\cG} &=&K_{0}\Big[\prod_{v \in \cV} (-\lambda_{v})\Big] \int \Big[\prod_{v \in \cV} \md p_{0;v} \prod_{s}\md p_{v, s}\Big] 
\Big[\prod_{\ell \in \cL} C_{\ell}
(\{p_{0;v(\ell)}, \vec p_{v(\ell), s}\};\{p'_{0;v'(\ell)}, \vec p\,'_{v'(\ell), s}\})\Big]  
\cr\cr
&\times&\Big[
\prod_{v\in \cV }\prod_{s}\delta(p_{v, s}-p'_{v, s'}) \Big]
 \Big[\prod_{v\in\cV} \delta(\sum_{l=1}^4 p_{0;l;v}) \Big]\,,
\eea
where $\lambda_v$ is a given coupling associated with $v \in \cV$;  
$p_{0;\ell}$ and $p_{v(\ell), s}$ are the coordinates involved
in the propagator labelled by a line index  $\ell$ incident
to its source and target vertices $v(\ell)$ and $v'(\ell)$; $p_{v,s}$ are the $p$ coordinates of the vertex $v$ and they  possess a strand index $s$. 
The constant $K_0$ includes  the Fermionic Pfaffian signs, the graph symmetry factor and a
combinatorial constant. We will use the compact
notation $K_{0}\Big[\prod_{v \in \cV} (-\lambda_{v})\Big]=\kappa(\lambda)$.

Note that the propagator in the $\vec p$ coordinates
is a product of Kroneckers $\delta$ and, similarly, the vertex kernels are also 
products of $\delta$ functions which convolute the different indices of
the tensors. Integrating those $\delta$
produces conservation of the $p$ coordinate index along 
a strand of the tensor graph. 
At the end of integration of all $\delta$'s in all propagators,
one obtains a $p_f$ coordinate per one-dimensional object  $f$ in the graph
that we call \emph{face}. 
Graphically a face is an alternating sequence of propagator lines with color $0$ and
colored lines $c$ of vertices. A face is \emph{closed} or \emph{internal} if this sequence is a  cycle in the colored  graph; it is otherwise \emph{open} or \emph{external}.   
 The set of closed faces is denoted $\cF_{\inter}$, and its cardinality $F_{\inter}$; the set of open faces is denoted $\cF_{\ext}$, 
and its cardinality $F_{\ext}$. We write $\cF_{\inter}\cup \cF_{\ext} =\cF$
the set of all types of faces. 
Given a closed face (resp. open face) $f$, we denote $p_f$ (resp. $p^{\ext}_{f}$) the 
 momentum coordinate associated with $f$.

A face $f$ is made of lines, hence we write $\ell \in f$. 
We introduce an incidence matrix between line and faces which identifies
if a line goes through a face or not: 
\bea
\epsilon_{\ell f} = \left\{ \begin{array}{ll} 1, &  \text{ if } \ell \in f \\
0, & \text{ otherwise }
\end{array} \right. 
\eea
We expand the amplitude \eqref{ampl0} 
as follows: 
\bea\label{ampl1}
A_{\cG} &=& \kappa(\lambda) 
 \int [\prod_{\ell\in \cL}\md\alpha_{\ell} ]
\int [\prod_{v \in \cV} \md p_{0;v}] [ \prod_{f \in \cF_{\inter}}\md p_{f}]
\Big[[\prod_{\ell\in \cL_1}  (-ip_{0;\ell})][ \prod_{\ell \in \cL_2}
 e( \sum_{f\in \cF}\epsilon_{\ell f} p_f) ]\Big]  \cr\cr
 & \times  &
 e^{-\alpha_{\ell}( p_{0;\ell}^2 +e^2( \sum_{f\in \cF}\epsilon_{\ell f} p_f) ) }\Big]
 \Big[\prod_{v\in \cV} \delta(\sum_{l=1}^4p_{0;l;v})\Big] .
\eea
This amplitude must be regularized by a cut-off on momenta
from which we will able to discuss the behavior of that amplitude. 
This is the task of the next section.

\section{Multi-scale Analysis and Power Counting}

We obtain, in this section, a power counting theorem 
for the amplitudes  \eqref{ampl1} in the ultraviolet regime
using a multi-scale analysis of the Feynman amplitudes in the spirit 
of \cite{Rivasseau:1991ub}, adapted to the tensor context of nonlocal actions. 

We begin with the slice decomposition of the propagator. This is a
decomposition of the parametric integral  obtained from \eqref{propagChi}
using a  geometric progression with ratio $M>0$. We write 
\bea
&&
\hat C_{\s\s'}(p_0, \vec p) = \sum_{i=1}^\infty C_{\s\s';\;i}  (p_0, \vec p)\,, 
\crcr
&&
C_{\s\s';\;i}  (p_0,\vec p) =  \int_{M^{-2i}}^{M^{-2(i-1)}} \md\alpha \, (-ip_0\delta_{\s\s'} +\varepsilon_{\s\s'} e(p) )  e^{-\alpha  (p^2_0  + e^2(p))}
\label{slice}
\eea
where we introduce the anti-symmetric tensor $\varepsilon_{\s\s'}$,
such that $\varepsilon_{12}=-1$ according to \eqref{propagChi}. 

An ultraviolet cut-off is imposed in the space of indices $i$, such that
$C^{\rho} = \sum_{i=1}^\rho C_i$ is the 
cut-offed propagator. The ultraviolet-limit is obtained by taking $\rho \to 
\infty$. We omit to write the symbol $\rho$ on each propagator, for simplicity. 
We expect our theory to be fully consistent in the $\rho\to \infty$ limit but this issue is postponed to a future study. In this paper we only establish 
perturbative renormalization at all orders. 

In this ultraviolet regime the value of the chemical potential is unimportant and we have the rather trivial bound
\bea\label{bounds}
\vert C_{\s\s';\; i}  (p_0,\vec p)\vert  &\leq&  K M^{-i}  \;
e^{- \delta M^{-i}( \vert p_0 \vert  + p^2)} \,,
\eea
for some constants $K$ and $\delta$. Remark the anisotropy between $p_0$
and $p$ and the fact that this bound does not depend on the $\s$ indices. We therefore 
simplify our notations and omit to mention these in the remaining analysis. 

The multiscale analysis allows for an optimal amplitude bound.
We consider a connected amputated Feynman graph $\cG$ of the theory
 with vertex set $\cV=\cV_+\cup \cV_m$, $V=|\cV|$
with propagator line set $\cL$, $L= |\cL|$. 
We work at this stage with amputated amplitudes, that
are graphs with external vertices where test  functions or external fields can be inserted.  The number of those external fields also called 
external legs is $N_{\ext}$.
 
Introduce the multi-index $\bmu \in \mathbb{N}^{L}$  
called (index) assignment which gives to each propagator 
 line $\ell$ of the graph  a scale $i_\ell\in [\![0,\rho]\!]$. 
Slicing all propagators, the initial amplitude becomes 
$A_{\cG} =\sum_{\bmu}A_{\cG;\bmu}$ where 
$A_{\cG;\bmu}$ is called the multi-scale representation of 
the amplitude $A_{\cG}$. 
After renormalizing the theory, 
the sum over $\bmu$ or over all possible assignments
will be performed. We have at fixed index assignment $\bmu$: 
\bea\label{amu}
A_{\cG;\bmu} &=&
\kappa(\lambda) \int [\prod_{v \in \cV} \md p_{0;v} \prod_{s}\md p_{v, s}] 
[\prod_{\ell \in \cL} C_{i_\ell}
(\{p_{0;v(\ell)}, \vec p_{v(\ell), s}\};\{p'_{0;v'(\ell)}, {\vec p}\,'_{v'(\ell), s}\})] 
\cr\cr
&\times& \Big[
\prod_{v\in \cV} 
\prod_{s}\delta(p_{v, s}-p'_{v, s'})\Big]
\Big[\prod_{v \in \cV} \delta(\sum_l p_{0;l;v})\Big]
\;.
\eea

Our goal is to find an optimal bound on $A_{\cG;\bmu}$  using, as much as possible, 
the decay of the lines. To do so, we introduce
the so-called quasi-local subgraphs $\cG^{i}$ of $\cG$
as the subgraphs made of lines of $\cG$ with index higher 
than $i$: $\forall \ell \in \cL(\cG^i)\cap \cL, i_\ell \geq i$. 
$\cG^i$ might have several connected components that
we denote at fixed $i$, $\cG^{i}_{(k)}$. Then $\{\cG^i_{(k)}\}$
is the set of all quasi-local subgraphs of $\cG$. 
Consider $g$ a subgraph of $\cG$, seeking a criterion for checking
if $g$ should coincide with some $\cG^i_{(k)}$, we have the following: 
at fixed index assignment $\bmu$, define 
\beq
i_g(\bmu)=\inf_{l \text{ internal line } \in g} i_l \qquad 
\text{ and }  \qquad 
e_g(\bmu)=\sup_{l \text{ external line }\in g}i_l\,,
\eeq
then  there exists $(i,k)$, such that $g =\cG^i_{(k)}$ if and only if 
$ i_g(\bmu) >e_{g}(\bmu)$, the so-called almost local condition. 
The value of $i$ satisfies $ i_g(\bmu)\ge i >e_{g}(\bmu)$. 
An important property of the set of quasi-local graphs $\{\cG^i_{(k)}\}$ is that it is partially ordered under inclusion,
and, using this partial order, one forms an abstract tree
namely the Gallavotti-Nicol\'o (GN) tree \cite{Galla}. The rest of our program
is to find an optimal bound for $A_{\cG;\bmu}$ in terms of the nodes
of the GN tree, in other words, an optimal bound
which expresses uniquely in terms of the $\{\cG^i_{(k)}\}$. 

At a fixed scale index $i$, we will need the following approximation of the sum  
\be\label{}
\sum_{p\in \Z} e^{- \delta M^{-i} |p|^{n} }
 = c M^{\frac{i}{n}}(1+ O(M^{-\frac{i}{n}}))
= \int_{-\infty}^{\infty} \md p \, e^{- \delta M^{-i} |p|^{n} }\,,
\ee
for  $n>0$ and some positive constant $c$
(see the detail of  the calculations in Appendix A of \cite{Geloun:2013saa}). 

We are ready to perform the integration over internal variables of the $\{\cG^i_{(k)}\}$ graphs.
This can be organized in completely equivalent ways either in momentum or direct space, using respectively the bound \eqref{bounds}, the important point being that it has to follow the GN tree structure. 
It means
we sum inductively over the internal $(p_0,\vec p)$ loop momenta of the $\{\cG^i_{(k)}\}$ graphs,
following the GN tree structure. At fixed $\bmu$,  we integrate over delta's in the $p$-space 
and use \eqref{bounds} to obtain
\bea\label{ampl2}
|A_{\cG;\bmu}| &\leq&  
K_1 \Big[\prod_{\ell\in \cL}M^{-i_{\ell}} \Big]
\int [\prod_{v \in \cV} \md p_{0;v}]
\Big[\prod_{\ell \in \cL}
e^{- \delta M^{-i_{\ell}} |p_{0;\ell}| }   \Big] \Big[\prod_{v\in \cV}\delta(\sum_l p_{0;l;v})\Big]  \crcr
& \times & \int [\prod_{f\in \cF_{\inter}}\md p_{f}]  \Big[
\prod_{f\in \cF_{\inter}}
e^{-  \delta (\sum_{\ell \in f} M^{-i_{\ell}}) p^2_{f} } \Big]  \,,
\eea
where $K_1= K^L \kappa(\lambda) K_{\ext}$, and $K_{\ext}$
is a bound over the product of external face amplitudes  
$e^{-  \delta (\sum_{\ell \in f} M^{-i_{\ell}}) p^2_{\ext; f} } $
which can be easily achieved by bounding each factor by a constant. 
Note that the r.h.s bound factorizes along $p_0$-space and $p$-space.
To find an optimal bound amplitude is therefore like combining
a standard local QFT procedure and a nonlocal one.  

The goal is to make the result of that summation/integration 
 as low as possible. The integration over $p_{0;l;v}$ variables is standard in ordinary local QFT:
we choose a vertex root and perform a momentum routine over the  $p_{0;l;v}$.
We can integrate over the set  ${\rm Cycle}_{\cG}$ of independent cycles (loops in the underlying graph); along each cycle $c$ choose the minimal index among the $i_\ell$'s: $i_c =\min_{\ell \in c}i_\ell$.  In direct space, this is choosing a tree   compatible with the GN tree, as explained in \cite{Rivasseau:1991ub}. 
Concerning the nonlocal part, for each internal face $f$, we introduce the index 
$i_f = \min_{\ell \in f}i_\ell$ that will be important during the integration. 

We are in position to find an optimal
bound for any amplitude \eqref{ampl2} as
\bea\label{ampl3}
|A_{\cG;\bmu}| \leq  
K_1
K_2\Big[ \prod_{\ell\in \cL}M^{-i_{\ell}} \Big]
\Big[\prod_{c \in {\rm Cycle}_{\cG}}
M^{i_c} \Big] 
\Big[\prod_{f\in \cF_{\inter}}
M^{\frac{i_f }{2} } \Big] 
\eea
where $K_2$ is a constant. 
This result must be re-expressed in terms of the quasi-local subgraphs. 
Each of the factors have been already addressed in previous works. 
We have
\bea\label{ampl4}
&&
|A_{\cG;\bmu}| \leq  
K_3\Big[ \prod_{\ell \in \cL} \prod_{i=1}^{i_\ell}M^{-1} \Big]
\Big[\prod_{c \in {\rm Cycle}_{\cG}}
\prod_{i=1}^{i_c}
M\Big] 
\Big[\prod_{f\in \cF_{\inter}}
\prod_{i=1}^{i_f}
M^{\frac{1}{2} } \Big] \\
&&
\leq  
K_3\Big[ \prod_{\ell \in \cL} \prod_{(i,k)/ \ell \in \cL(\cG^{i}_{(k)})} M^{-1} \Big]
\Big[\prod_{c \in {\rm Cycle}_{\cG}}
\prod_{(i,k)/ \ell \in \cL(\cG^{i}_{(k)})}
M \Big] 
\Big[\prod_{f\in \cF_{\inter}}
\prod_{(i,k)/\ell \in \cL(\cG^{i_f}_{(k)})}
M^{\frac{1}{2}  } \Big] 
\nonumber
\eea
with $K_3 = K_1K_2$.  
The two first products are well-known (see \cite{Rivasseau:1991ub}) and we simply rewrite them as: 
\bea
&&
\prod_{\ell \in \cL} \prod_{(i,k)/ \ell \in \cL(\cG^{i}_{(k)})}M^{-1}  = \prod_{(i,k)} M^{-L(\cG^i_{(k)})}\,, 
\crcr
&&
\prod_{c \in {\rm Cycle}_{\cG}}
\prod_{(i,k)/ \ell \in \cL(\cG^{i_c}_{(k)})}
M 
 = \prod_{(i,k)}M^{[L(\cG^i_{(k)})-(V(\cG^i_{(k)}) -1)]} 
 \label{cycl}
\eea
The last product  decomposes into the following 
\be
\prod_{f\in \cF_{\inter}}
\prod_{(i,k)/\ell \in \cL(\cG^{i_f}_{(k)})}
M^{\frac{1}{2}  } 
= \prod_{f\in \cF_{\inter}}
 \prod_{(i,k)/\ell_{f} \in \cL(\cG^i_{(k)})}
M^{\frac{1}{2}  } 
 =  \prod_{(i,k)} \prod_{f\in \cF_{\inter} \cap \cG^i_{(k)}}
M^{\frac{1}{2}  }  = \prod_{(i,k)}M^{\frac{1}{2} F_{\inter}(\cG^i_{(k)}) }
\label{fintern}
\ee
Let us address now the vertices coming from the mass and wave functions
couplings. The introduction of a mass coupling does not change
the overall analysis. Adding wave function vertices  changes the power counting
by introducing an equal number of vertices $V_{p_0}$ and $V_{p^2}$. 
The above analysis leading to \eqref{cycl} remains the same, 
the exponent therein becomes 
\be
L(\cG^i_{(k)})-( V(\cG^i_{(k)}) +V_{p_0}(\cG^i_{(k)}) 
+V_{p^2}(\cG^i_{(k)})  -1)\,, 
\ee
where $ V(\cG^i_{(k)})$ uniquely denotes the number of quartic vertices. 
Then the vertex weights $(-ip_0)$ and $p^2$ introduce back a factor
\be
\prod_{(i,k)} M^{V_{p_0}(\cG^i_{(k)}) + V_{p^2}(\cG^i_{(k)})  }
\ee
Collecting all contributions, the following statement holds: 
\begin{theorem}[Power counting]
\label{theo:powcount}
Let $\cG$ be a connected graph of the model \eqref{act:tens}
with Gaussian measure determined by 
the covariance \eqref{propagChi}.
Considering $A_{\cG;\bmu}$ the amplitude associated with $\cG$ 
at  index assignment $\bmu$, there exists some large constant $K$ such that
\beq\label{divdeg}
|A_{\cG;\bmu}| \leq  K^{V(\cG)} \prod_{(i,k)\in \mathbb{N}^2} M^{\omega_{\deg}(\cG^i_{(k)})} \,, \
\eeq
where $\cG^i_{(k)}$ are the quasi-local subgraphs and
the divergence degree is given by
\be\label{degrediver}
\omega_{\deg}(\cG) = - (V(\cG)-1)
+ \frac{1}{2} F_{\inter}(\cG)  \,. 
\ee
\end{theorem}

From the above theorem, we clearly see that the model that will be renormalizable
is of a different type than usual scalar field theory or usual tensor field theory.

\section{Analysis of the divergence degree}
\label{sect:potent}

We need to 
count the number of internal faces in a graph $\cG$ \emph{with external legs}.
This requires to extend the notion of jackets into \emph{pinched jackets} \cite{BGR}.
This is usually done in a bipartite (complex) framework
but in our case we have real fields and the graphs are not bipartite
so we shall use some new arguments.

\begin{proposition}\label{Fint}
Consider a connected rank $d=3$ graph $\cG$, with 
boundary graph $\bG$. Let $C_{\partial}$ be the number of
connected components of $\bG$, $V_+$ the number
of vertices of the kind $\bb_+$ and $V_m$ the
number of vertices of the kind $\bb_c$
$V=V_+ + V_m $.
$N_{\ext}$ is the number of external legs of $\cG$; 
\be\label{facinter}
F_{\inter}(\cG)  = - (\omega(\cG_{\col}) -g_{\bG}  ) + 3V_+ + 2V_m  -  N_{\ext}  - (C_{\pa}-1) +2\,, 
\ee
where
$\omega(\cG_{\col}) = \sum_{\tJ}g_{\tJ}$ is the sum of genera of the pinched jackets of $\cG_{\col}$ the colored extension of $\cG$ and $g_{\bG}$ is the genus  of the boundary graph. 
\end{proposition}
\proof 
Consider $\cG$ a connected tensor graph and $\cG_{\col}$ the colored extension of $\cG$. 
We denote the number of vertices, $V_{\col}$, 
the number of lines, $L_{\col}$ of $\cG_{\col}$.
We recall our notations, $V$ and $L$ 
are, respectively, the same quantities for $\cG$, 
while $F_{\inter}$ is the number of internal faces of $\cG$. For the boundary graph $\bG$, we denote $V_{\partial}$, $E_{\partial}$ and
$F_{\partial}$ the cardinality of the vertex set, edge set
and face set. 
For a pinched jacket $\tJ$,  we use $V_{\tJ}$ for the number of vertices,  $E_{\tJ}$ for the number of edges, and $F_{\tJ}$ for the number (necessarily closed) faces. Note that because $\cG$ is connected,
so is $\cG_{\col}$ and any jacket within $\cG_{\col}$ is also
connected. 

There are $d!/2=3$ jackets in $\cG_{\col}$. Each 
$\bb_{c}$ or $\bb_+$ vertex of $\cG$ decomposes in 4 vertices  in $\cG_{\col}$.
Each of those vertices in $\cG_{\col}$ decomposes again in 3, 
for each of the jackets. 
Each line of $\cG$ splits in 3 to become an 
edge of a jacket.
Furthermore, each vertex $\bb_{c}$ or $\bb_+$ in $\cG$ is associated with 4 vertices in $\cG_{\col}$ which gives 6 additional colored lines.
This combinatorics gives:  
\be\label{edgvert}
\sum_{J} V_{\tJ} = 12 (V_++V_m)  \,,\qquad
\sum_{J} E_{\tJ} = 3 L + 18 (V_+ + V_m)\,. 
\ee
The number of faces of the pinched jacket $\tJ$ 
decomposes in 3 terms: 
\be
F_{\tJ} = F_{\inter; \, \tJ; \, \cG} +  F_{\inter;\, \tJ; \, \cG_{\col}}  + F_{\ext;\tJ} 
\ee
where $ F_{\inter;\tJ; \cG} $ is the number faces of $\tJ$ which belong 
to $\cG$ as well,   $F_{\inter;\tJ; \cG_{\col}} $ is the number of 
faces of $\tJ$ which belong to $\cG_{\col}$ but do not 
belong to $\cG$ and $F_{\ext;\tJ}$ are the faces of $\tJ$ which 
were external and are closed after pinching.

An internal face of a jacket contributing to $ F_{\inter; \, \tJ; \, \cG} 
+F_{\inter;\, \tJ; \, \cG_{\col}}$ is shared exactly  by another jacket; 
a jacket face contributing to $F_{\ext;\tJ} $ must be tracked 
at the level of the boundary graph $\bG$. 
We have, by summing over jackets:  
\be\label{intfaces}
\sum_{J} F_{\tJ} = 2 F_{\inter} + 2 F_{\col;\inter}  
+\sum_{J}  F_{\ext;\tJ} 
\ee
The quantity $F_{\col;\inter}$ is the number of additional internal 
faces brought by the colored expansion at the level of each 
vertex of $\cG$. Each vertex of the type $\bb_{c}$ brings
4 of such closed faces, meanwhile, a vertex of the cross type $\bb_{+}$
brings 3 of those.
This computes explicitly as
\be\label{colface}
F_{\col;\inter}= 3V_+ + 4V_m .
\ee
The last piece  of \eqref{intfaces} is now treated. Consider the boundary
graph $\bG$ which is a 3-regular ribbon graph. 
\be\label{boudve}
V_{\pa} = N_{\ext} \,, \qquad  
E_{\pa }  = F_{\ext} \,, \qquad 
3V_{\pa}   = 2 E_{\pa}\,. 
\ee
The boundary graph might have several connected components,
hence writing its Euler characteristic, we have
\be
2C_{\pa} - 2g_{\bG} = V_{\pa} - E_{\pa} +F_{\pa}  \,. 
\ee
Each face of $\bG$ can be uniquely mapped to a face 
of a unique pinched jacket which closes after pinching.  
Hence, 
\bea\label{bound}
\sum_{J} F_{\ext;\tJ} = F_{\pa} &=&  2C_{\pa} - 2g_{\bG}  - 
(V_{\pa} - E_{\pa}) \crcr
&=&
 2C_{\pa} - 2g_{\bG}  - (1- \frac32 ) N_{\ext} \crcr
&=&
2C_{\pa} - 2g_{\bG}  + \frac12 N_{\ext} \,. 
\eea
We are then in position to find an expression of the
number of internal faces of $\cG$. 
Combining the relations \eqref{edgvert}, \eqref{intfaces}, 
\eqref{colface} and \eqref{bound}, we get: 
\bea
F_{\inter} &=& \frac{1}{2}\Big[ \sum_{J} F_{\tJ} - 2 F_{\col;\inter}  
- \sum_{J}  F_{\ext;\tJ} \Big] \cr\cr
&=&
 \frac{1}{2}\Big[ \sum_{J}[2- 2g_{\tJ} - (V_{\tJ} - E_{\tJ} )] - 2 F_{\col;\inter}  
- \sum_{J}  F_{\ext;\tJ} \Big] \cr\cr
&=&
 \frac{1}{2}\Big[ 2\cdot 3 - 2\omega(\cG_{\col}) + 
3L +18(V_++V_m) - 12(V_++V_m) 
\crcr
 &-&   2 \Big[3V_+ + 4V_m \Big] 
- \Big[2C_{\pa} - 2g_{\bG}  + \frac12 N_{\ext} \Big]
\Big] \cr\cr
&=&- (\omega(\cG_{\col}) -g_{\bG}  ) + 3V_+ + 2V_m  
-  N_{\ext}  - (C_{\pa}-1) +2\,, \label{bigequa}
\eea
where we used the sum of the Euler characteristics
of connected pinched jacket $2-2g_{\tJ} = V_{\tJ} - E_{\tJ} - F_{\tJ}$,
define $\omega(\cG_{\col}) = \sum_{J}g_{\tJ}$  as
the degree of the graph, and use the relation $4(V_++V_m)
= 2L +N_{\ext}$. 

\qed 

\begin{proposition}[Divergence degree]\label{divdegr}
In the above notations,  
\be
\omega_{\deg} (\cG)
=-  \frac{1}{2}[\omega(\cG_{\col}) -V_+    - g_{\bG} +(C_{\pa}-1)]
-  \frac{1}{2} (N_{\ext}  -4) .  \label{omegadeg}
\ee
\end{proposition}
\proof We insert $F_{\inter}(\cG)$ of Proposition \ref{Fint} 
in \eqref{degrediver} of Theorem \ref{theo:powcount} and do some algebra
to obtain: 
\bea
\omega_{\deg} (\cG) &=&
- (V(\cG)-1)
+ \frac{1}{2} [- (\omega(\cG_{\col}) -g_{\bG}  ) + 3V_+ + 2V_m  
-  N_{\ext}  - (C_{\pa}-1) +2]    \cr\cr
&=&-   \frac{1}{2}[\omega(\cG_\col) -  V_+ -g_{\bG}  + (C_{\pa}-1)  ] -  \frac{1}{2} (N_{\ext}  -4) \,. 
\nonumber
\eea  
which is  \eqref{omegadeg}. 
\qed 

Lemma 7 in \cite{Gurau:2011tj} with $D=d=3$  states
that for vacuum graphs
\be
\omega(\cG_{\col}) \geq 3 \Big[\sum_{\bb_c} \omega(\bb_c) 
+ \sum_{\bb_+} \omega(\bb_+)\Big] .
\ee
 Using $\omega(\bb_c) = 0$ and  $\omega(\bb_+) = \frac{1}{2}$ 
we find
\bea
\omega(\cG_{\col}) \geq  \frac{3}{2}V_+  .
\eea
The quantity 
\bea \label{defindex}
\ind(\cG) = \omega(\cG_\col) -   \frac{3}{2} V_+ 
\eea
is called the index of the colored tensor graph $\cG$ \cite{FRV}. For vacuum graphs it coincides with the degree 
used in \cite{Carrozza:2015adg}.  Deleting lines in a vacuum graph 
can only decrease the genus hence even for graphs with external legs we have
\bea
\omega(\cG_{\col}) \geq  \frac{3}{2}V_+  \quad \Rightarrow \quad \ind(\cG) \geq 0\,.
\eea
In terms of this index
\bea
\omega_{\deg} (\cG)
=-   \frac{1}{2}[\ind (\cG) +  \frac{1}{2} V_+   -g_{\bG}  + (C_{\pa}-1)    ] -  \frac{1}{2} (N_{\ext}  -4) .  \label{omegaind}
\eea
From this point the renormalizability of the model could be addressed. 

\section{Renormalizability} 
\label{sect:renorm}

We now prove that the divergence degree is strictly negative for operators with 
6 or more external legs (also called convergent or irrelevant).
\begin{lemma}[Bound on convergent graphs with $N_{\ext} \ge 6$]\label{omGcol}
For $\cG$
any graph with $N_{\ext} \ge 6$,
\bea
\omega_{\deg} (\cG) \;\; {\le } \;\;- \frac{1}{12} N_{\ext} .
\eea
\end{lemma}
\proof
 Remark first that we need to prove the theorem only in the
case  $(C_{\pa}-1) =0 $  since 
considering disconnected boundaries  makes $\omega_{\deg}$  smaller.
In this case
\bea
\omega_{\deg} (\cG)
=-  \frac{1}{2}[\ind(\cG)  +  \frac{1}{2} V_+   - g_{\bG} ] -  \frac{1}{2} (N_{\ext}  -4) . \label{omegadeg4}
\eea
Since $\ind(\cG) \ge 0$ (positivity of the index) and $V_+ \ge 0$, we have 
\bea
\omega_{\deg} (\cG) \le  \frac{1}{2} [ g_{\bG}   -
 (N_{\ext} - 4  )].\label{omegadeg6}
\eea
It is easy to check that 
\bea  g_{\bG} \le \frac{N_{\ext}}{4} - \frac{1}{2} \label{genussmall}
\eea
since a three-colored graph, like $\bG$ is, has at least 3 faces. The above relation
is derived using \eqref{boudve}. Therefore
\bea
\omega_{\deg} (\cG) \le   \frac{1}{2} \big[  \frac{N_{\ext}}{4} - \frac{1}{2}   -
 (N_{\ext} - 4  ) \Big]   =   -\frac{3N_{\ext}}{8} + \frac{7}{4}  \,,
 \label{omegadeg7}
\eea
and the latter expression can be bounded by $-N_{\ext}/12$ whenever $N_{\ext}\ge 6$. 
\qed

It remains to treat the case of graphs with $N_{\ext}\leq 4$. 

\medskip
\noindent{\bf Four-point subgraphs.}
Let us set $N_{\ext}=4$, then by \eqref{omegadeg} the divergence degree
for these graphs is 
\be
\omega_{\deg} (\cG)
= - \frac{1}{2}[\ind(\cG)  +  \frac{1}{2} V_+   - g_{\bG} +  (C_{\pa}-1) ]  
  \label{omegadegN4}
\ee
and we want to check that $\omega_{\deg} (\cG) \le 0$ so that we have at most logarithmic divergence
for four point functions. Having four external legs, a graph 
can have three types of possible boundaries:

\begin{itemize}

\item A disconnected boundary, hence $\bG$ is made of two quadratic melons.
In that case $C_{\pa} = 2$ and $ g_{\bG} =0$ so that 
\be 
\omega_{\deg} (\cG)
= - \frac{1}{2}[\ind(\cG)  +  \frac{1}{2}V_+ +1 ]  \le - \frac{1}{2}[\ind(\cG)   +1 ] 
\; \le \; - \frac{1}{2}.
\ee
This case does not require renormalization.

\item A connected boundary with $\bG$ of the quartic melonic type  $\bb_c$, 
for some color $c$.
In that case $C_{\pa} = 1$, $ g_{\bG} =0$ so that 
\be 
\omega_{\deg} (\cG) = - \frac{1}{2} [\ind(\cG)  +  \frac{1}{2}V_+  ] \leq  0 
\ee
can be zero if $\ind(\cG)   =0= V_+$. 
 In particular, there is such a non-trivial graph at one loop, with $V_m= 2$. 
 This case certainly requires renormalization treatment. 

\item A connected boundary with $\bG$ of the $\bb_+$ 
type.
In that case $C_{\pa} = 1$ and $ g_{\bG} =\frac{1}{2}$, so that  
\be 
\omega_{\deg} (\cG) = - \frac{1}{2} [\ind(\cG)  +  \frac{1}{2} (V_+  - 1) ] .
\ee
The following subcases could be discussed: 

-  $V_+ >1$, then directly $\omega_{\deg} (\cG) <0 $, hence all this class define
graphs with convergent amplitude.

- $V_+=0$, this case is impossible to occur since the boundary is
non-orientable $g_{\bG}=\frac 12$, there must be some non-orientable
vertices.

- $V_+ =1$. This is the final and most delicate point. We obtain $\omega_{\deg} (\cG)  \le 0$, as expected. 
Apparently  the bound could saturate, namely $\omega_{\deg} (\cG)  =0$, 
when $\ind(\cG) = \omega(\cG_\col)-\frac 32=0$. But a more careful analysis 
shows that this is impossible. More precisely we shall prove 
\begin{lemma} If $N_{\ext}=4$, $V_+ =1$ and $g_{\bG}=\frac 12$, then $F_{\inter} \le 2V_m -1$, hence 
by \eqref{bigequa} and \eqref{omegadeg} $\omega_{\deg} (\cG) \le - \frac{1}{2}$.
\end{lemma}
\proof
Let us call $\cG '$ the graph made from $\cG$ by cutting out $V_+$. It has $V_m$ vertices, all of melonic type.
The case  $V_m=1$ is easy, as $F_{\inter} =1$ in that case. Then 
we can complete the proof that $F_{\inter} \le 2V_m -1$ by induction. 
If the vertex $V_+$ is attached to 2 external lines,  $\cG'$ is made of melonic vertices and has $4$ external legs, hence
its number of faces is maximal if $\cG'$ is \emph{fully melonic}, in which case it has
$2(V_m -1)$ internal faces (the melonic rate). Joining $\cG'$ to $V_+$ creates at most one new internal face and we are done.

If the vertex $V_+$ is attached to 2 external lines, since
$\cG'$ has $6$ external legs, 
it can have \emph{at most $2(V_m -2)$ internal faces (again the maximal melonic rate)}. 
Joining $\cG'$  to $V_+$ creates at most three new internal faces and we are done again.

Finally when the vertex $V_+$ is attached to no external lines,
$\cG'$  has $8$ external legs, hence
at most $2(V_m -3)$ internal faces (again the melonic rate). Joining $\cG'$ to $V_+$ can 
creates at most six new internal faces, hence we are not done yet. To gain the crucial last improvement of one face, we shall prove  that in this case the boundary graph cannot be of the $V_+$ type. Indeed if $\cG'$ has exactly $2(V_m -3)$ internal faces,
its boundary must be a melonic colored graph with eight vertices. But if this graph, when joined to $V_+$, creates 6 
additional faces, it must be that its boundary was \emph{disconnected} into at least two pieces with 4 colored vertices each
(since all circuits of the four external legs of $\cG'$ joined to $V_+$ have to be internal). Under that condition of disconnected boundary the maximal ``melonic'' number of internal faces is no longer  $2(V_m -3)$ but $2(V_m -2)$ and we are done.
\qed

\end{itemize}

\medskip
\noindent{\bf Two-point subgraphs.}
There is no longer any choice for the boundary, as $g_{\bG} = 0$ and $C_{\pa}=1$ (there is only a single invariant with two vertices).
The degree of divergence takes the form: 
\be
\omega_{\deg} (\cG)
=-  \frac{1}{2} [\ind(\cG)  +  \frac{1}{2} V_+    ] 
 + 1  =  -  \frac{1}{2} [\omega(\cG_\col)   -   V_+    ] 
 + 1 \label{omn2}
\ee
and is at most 1.  As usual this means that we should perform mass and wave-function subtractions.
We have therefore $\omega_{\deg} (\cG) \ge 0$ equivalent to  $ \omega(\cG_\col)   -   V_+   \in \{0,1,2\}$.

To summarize we have proved
\begin{theorem}
\noindent$\bullet$ If $N_{\ext} \ge 6$
\be 
\omega_{\deg} (\cG) \le - \frac{N_\ext}{12}
\ee
hence these functions are convergent.

\noindent$\bullet$ If $N_{\ext} =4$
\be 
\omega_{\deg} (\cG) \le 0
\ee
hence four-point functions are at most log-divergent and renormalized by a single subtraction.

\noindent$\bullet$ If $N_{\ext} =2$
\be 
\omega_{\deg} (\cG) \le 1
\ee
hence two-point functions are at most quadratically divergent.
\end{theorem}

\section{Renormalization}
\label{sec:renorm}

This section undertakes the renormalization of the divergent graphs of the model. 
We focus on the expansion of the amplitudes around their divergent and ``local'' 
part. The goal is to subtract the  local part of quasi local graphs and this improves power counting
of the amplitudes. 

There are two types of graphs which have divergences: four- and two-point diagrams. 
They will be treated separately. 

We consider amplitudes with external legs. There are therefore two types of lines 
in a diagram, internal lines that we denote $l$ and external lines denoted $l_\ext$. 
An internal line $l$ is associated with a high scale $i_l$ of an internal momentum and 
   a parameter $\alpha_l \in [M^{-2i_l}, M^{-2(i_l-1)}]$. 
An external line $l_\ext$ is associated with a lower scale $j_{l_\ext} < i_l $ of an external momentum, 
and a parameter $\alpha_{l_\ext} \in [M^{-2j_{l_\ext}}, M^{-2(j_{l_\ext}-1)}]$. 

We have two types of momenta: \emph{time} momenta $p_{0}$ 
and \emph{space} momenta $p$. Their treatment in the following expansion
is different and urge us to introduce more notations. 
For space momenta,  $p^{\ext}_f$ is associated with an external face $f$ and
and $p_f$ denotes an internal momenta associated with  a closed face. 
External time momenta associated with external lines are denoted $p_{0;l_\ext}$ 
and those associated internal lines are denoted by $p_{0;l}$. Note that, 
since there is conservation of time momenta at the vertices, the $p_{0;l}$'s
might be very well  (linearly) depending on $p_{0;l_\ext}$. 
After imposing the vertex constraints, it remains one internal 
momenta per independent cycle $c$ in the graph.  

\medskip
\noindent{\bf Four-point amplitudes.} 
Consider a four-point function which is log-divergent. 
It is of the boundary type: $g_{\bG} =0$.
Pick a diagram amplitude coming from the expansion of the 
correlator: 
\be\label{corr}
\langle \chi_{1;\;p_{0;1} 123} \,\chi_{2;\;p_{0;2}1'23}\, \chi_{1;\;p_{0;3}1'2'3'}\, \chi_{2;\;p_{0;4} 12'3'} \rangle
\ee
where the notation $ \chi_{\s;\;p_{0;i} 123}$  stands for  $ \chi(p_{0;i},p_1,p_2,p_3,\s)$. Note that this correlator
has a boundary graph which is of the form of the melonic interaction with particular color 1
(this is the bubble $\bb_1$). 
We will perform the expansion of an amplitude with this boundary data, to perform a similar analysis
for other melonic boundary with color $c=2,3$ will be straightforward. 

We start by noting that a diagram issued from \eqref{corr} has four external 
propagator lines  with momenta $p_{0;a}$, $a=1,2,3,4$, that we associate
with  external lines $l_{\ext}$ (depending of course on $a$) such that  $p_{0;a}= p_{0;l_\ext}$. 

A graph amplitude of the model is of the form  
\bea\label{ampl41}
&&
A_{\cG;4}(\{p_{0;l_\ext}\};\{p^{\ext}_f\}) = \kappa(\lambda) 
 \int [\prod_{\ell\in \cL}\md\alpha_{\ell} ]
\int [\prod_{v \in \cV} \md p_{0;v}] [ \prod_{f \in \cF_{\inter}}\md p_{f}]
\Big[[\prod_{\ell\in \cL_1}  (-ip_{0;\ell}][\prod_{\ell\in \cL_2}  e( \sum_{f\in \cF}\epsilon_{\ell f} p_f) ]\Big] \cr\cr
&&
\times  e^{-\sum_{\ell\in \cL} \alpha_{\ell}\, e^2( \sum_{f\in \cF}\epsilon_{\ell f} p_f) }\;
  e^{-\sum_{\ell\in \cL} \alpha_{\ell} \,p_{0;\ell}^2} \Big[\prod_{v\in \cV} \delta(\sum_{l=1}^4p_{0;l;v})\Big] \cr\cr
 &&=\kappa(\lambda) 
 \int [\prod_{\ell\in \cL}\md\alpha_{\ell}  e^{\alpha_\ell}]
\int [\prod_{v \in \cV} \md p_{0;v}] [ \prod_{f \in \cF_{\inter}}\md p_{f}]
\Big[[\prod_{\ell\in \cL_1}  (-ip_{0;\ell}][\prod_{\ell\in \cL_2}  e( \sum_{f\in \cF}\epsilon_{\ell f} p_f) ]\Big]  \cr\cr
&&
\times 
[ \prod_{f \in \cF_{\ext} } e^{-(\sum_{\ell \in f}\alpha_\ell) [(p^\ext_f)^{4} -2 (p^\ext_f)^{2}] }] 
[ \prod_{\substack{f,f' \in \cF_{\ext}\\ f \ne f' }}e^{-(\sum_{\ell \in f, \ell \in f'}\alpha_\ell) (p^\ext_f)^{2} (p^\ext_{f'})^{2} } ]\cr\cr
&& \times 
[  \prod_{ f \in \cF_{\inter} } e^{-(\sum_{\ell\in f} \alpha_{\ell})[ p^4_f - 2p^2_f ] }]
[ \prod_{\substack{f,f' \in \cF_{\inter} \\ f \ne f'}}e^{-(\sum_{\ell \in f, \ell \in f'}\alpha_\ell) (p_f)^{2} (p_{f'})^{2} } ]
\cr\cr
&& \times
[ \prod_{f \in \cF_{\ext}, f' \in \cF_{\inter} }e^{-2(\sum_{\ell \in f, \ell \in f'}\alpha_\ell) (p^\ext_f)^{2} (p_{f'})^{2} } ]
\times   e^{-\sum_{\ell\in \cL} \alpha_{\ell} \,p_{0;\ell}^2} \Big[\prod_{v\in \cV} \delta(\sum_{l=1}^4p_{0;l;v})\Big] \,.
\eea 
Consider the decomposition of the set $\cL$ of lines in internal lines $\cL_{\inter}$
and external lines $\cL_{\ext}$. The treatment of the  momenta $p_{0;\ell}$ resorts from a usual technique: first, lines must be oriented 
in an arbritrary way (but, at the end, the procedure is independent of the orientation);
at  each vertex $v$, if a line $l$ is oriented towards $v$, 
the sign of the momentum $p_{0;l}$ associated with $l$ in the $\delta$-function is chosen positive,
 and negative otherwise; second we must fix a tree
 $\cT$  of internal lines and do a momentum routine along the lines of that tree. 
 Using the $\delta$-functions of the vertices, and expanding the squares produces
 a sign before the Schwinger parameter $\alpha$, that we denote $\al= \pm \alpha$. The following development and the conclusion of our analysis do not actually depend on the signs and we will keep a general notation $\al$ without a concern about these signs. 
Note that for a given cycle $c\in {\rm Cycle}_{\cG}$ of the graph that corresponds to a given high momentum $p_{0;c}$,  there is a subset  $\cT_c \subset \cT$  of lines. 
There is  a line $l_c \in \cL_{\inter}$ such that  the set of lines $\{l_c\} \cup \cT_c = \cL_c$ forms the cycle $c$. 
 With  each external momenta $p_{0;l_\ext}$, there is a path $\cT_{l_\ext}\subset \cT$ 
of internal lines $l$ such that after the integration of the $\delta$-functions, $p_{0;l}$
becomes a function of $p_{0;l_\ext}$. We then introduce another matrix, $|\cL_{\inter}|\times 
(|{\rm Cycle}_\cG|+|\cL_{\ext}|)$,  which decomposes
in to diagonal blocks: 
\be
\varepsilon_{lc} = \left\{\begin{array}{ll} 1 & \text{if}\;\; l \in\cL_c   \\
0 & \text{otherwise} \end{array} \right. \qquad \qquad 
\varepsilon_{ll_\ext} = \left\{\begin{array}{ll} 1 & \text{if}\;\; l \in \cT_{l_\ext}  \\
0 & \text{otherwise} \end{array} \right. 
\ee

Then,  we have the following expansion:
\bea\label{expL}
&&
[\prod_{\ell\in \cL } e^{- \alpha_{\ell} p_{0;\ell}^2 } ][\prod_{v\in \cV} \delta(\sum_{l=1}^4p_{0;l;v})]  = 
  [\prod_{v\in \cV} \delta(\sum_{l=1}^4 p_{0;l;v})] [\prod_{ c \in {\rm Cycle}_{\cG} }
 e^{- (\sum_{l \in \cL_c } \alpha_l)  p_{0;c}^2 } ] [ \prod_{\substack{c,c' \in {\rm Cycle}_{\cG} \\ c\ne c'}}
 e^{- (\sum_{ l \in \cT_c\cap \cT_{c'}  } \al_{l})p_{0;c}p_{0;c'} } ]
\crcr
 &&
 \times 
[\prod_{l_\ext\in \cL_{\ext} }
 e^{- (\alpha_{l_\ext} +\sum_{ l \in  \cT_{l_\ext} } \alpha_{l}) p_{0;l_{\ext}}^2  } ]
[ \prod_{\substack { l_\ext\in \cL_{\ext}\\ c \in  {\rm Cycle}_{\cG}}}
 e^{- 2(\sum_{l \in \cT_c\cap \cT_{l_\ext}  } \al_{l}) p_{0;c} p_{0;l_{\ext}} } ]\cr\cr
 && \times 
[ \prod_{\substack{l_\ext,l_{\ext}'\in \cL_{\ext}\\ l_\ext\ne l_{\ext}'} }
 e^{- (\sum_{ l \in \cT_{l_\ext}\cap \cT_{l'_\ext}  } \al_{l})p_{0;l_{\ext}}p_{0;l_{\ext}'} } ]\,, 
\eea
where we used the $\delta$-functions to perform the relevant substitutions. 

We perform the following expansion for each factor associated
with external momenta: 
\bea
&&
 e^{- (\alpha_{l_\ext} +\sum_{ l \in  \cT_{l_\ext} } \alpha_{l}) p_{0;l_{\ext}}^2  } 
  =
 e^{- \alpha_{l_\ext} p_{0;l_{\ext}}^2  } 
 \Big[  1- Q^{1}_{l_\ext}\Big] \cr\cr
 &&
 Q^1_{l_\ext} = 
 (\sum_{ l \in  \cT_{l_\ext} } \alpha_{l}) \, p_{0;l_{\ext}}^2 
 \int_0^1 \md s\,
 e^{-s (\sum_{ l \in  \cT_{l_\ext} } \alpha_{l}) p_{0;l_{\ext}}^2  }\,,  \cr\cr
 && 
 e^{- 2 (\sum_{c \in  {\rm Cycle}_{\cG}}(\sum_{l \in \cT_c\cap \cT_{l_\ext}  } \al_{l}) p_{0;c}) p_{0;l_{\ext}} } 
  = 1 -  Q^{2}_{l_\ext} \cr\cr
 && 
Q^{2}_{l_\ext}   =2
\Big[\sum_{c \in  {\rm Cycle}_{\cG}}(\sum_{l \in \cT_c\cap \cT_{l_\ext}  } \al_{l}) p_{0;c}\Big] p_{0;l_{\ext}}
 \int_0^1 \md s\,
 e^{- 2 s(\sum_{c \in  {\rm Cycle}_{\cG}}(\sum_{l \in \cT_c\cap \cT_{l_\ext}  } \al_{l}) p_{0;c}) p_{0;l_{\ext}} } \,, \cr\cr
 &&
 e^{- (\sum_{ l \in \cT_{l_\ext}\cap \cT_{l'_\ext}  } \al_{l})p_{0;l_{\ext}}p_{0;l_{\ext}'} } 
  =  1 -  Q^{3}_{l_\ext,l'_\ext} \cr\cr
  &&
  Q^{3}_{l_\ext, l'_{\ext}} 
   = (\sum_{ l \in \cT_{l_\ext}\cap \cT_{l'_\ext}  } \al_{l})p_{0;l_{\ext}}p_{0;l_{\ext}'}
    \int_0^1 \md s\,
    e^{- s(\sum_{ l \in \cT_{l_\ext}\cap \cT_{l'_\ext}  } \al_{l})p_{0;l_{\ext}}p_{0;l_{\ext}'} } \,. 
    \label{expl0}
\eea
Focusing on the momentum associated with space coordinates,  
for the momenta $p_{f}^\ext$ associated with an external face $f$, we use the following
decomposition and expansion: 
\bea
&& 
e^{-(\sum_{\ell \in f}\alpha_\ell) [(p^\ext_f)^{4} -2 (p^\ext_f)^{2}] }
  =  e^{  -(\alpha_{ l_{\ext} } + \alpha_{ l_{\ext}' }) [(p^\ext_f)^{4} -2 (p^\ext_f)^{2}]   }
   (1- Q^1_{\ext ;f})
 \cr\cr
 &&  
 Q^1_{\ext ;f} =     (\sum_{l \in f } \alpha_{l})  [(p^\ext_f)^{4} -2 (p^\ext_f)^{2}] 
   \int_0^{1} \md s\, e^{-s(\sum_{l \in f}\alpha_l )(p_f^{\ext})^2 }  \,,    \cr\cr
   && 
 e^{-(\sum_{\ell \in f, \ell \in f'}\alpha_\ell) (p^\ext_f)^{2} (p^\ext_{f'})^{2} } 
 =  e^{-(\alpha_{l_\ext}+\alpha_{l_\ext'}) (p^\ext_f)^{2} (p^\ext_{f'})^{2} }    (1- Q^2_{\ext ;f})
 \cr\cr
 && 
 Q^2_{\ext;f,f'} =     ( \sum_{l \in f, l\in f' } \alpha_{l})  (p^\ext_f)^{2}(p^\ext_{f'})^{2}
   \int_0^{1} \md s\,  e^{-s(\sum_{l \in f, l\in f'}\alpha_l )(p_f^{\ext})^2(p_{f'}^{\ext})^2 }  \,, 
   \cr\cr
   && 
  e^{-(\sum_{\ell \in f, \ell \in f'}\alpha_\ell) (p^\ext_f)^{2} (p_{f'})^{2} } 
  = 1-Q^3_{\ext;f,f'} \cr\cr
  &&
  Q^3_{\ext;f,f'} =   ( \sum_{l \in f, l\in f' } \alpha_{l})  (p^\ext_f)^{2}(p_{f'})^{2}
   \int_0^{1}\md s\, e^{-s(\sum_{l \in f, l\in f'}\alpha_l )(p_f^{\ext})^2(p_{f'})^2 }\,,
       \label{expf}
  \eea
where, in the last expansion, we use the fact that there is no external lines which could belong to $f'\in \cF_{\inter}$. 

It remains the following factor to study, for $l\in \cL_{\inter}\cap \cL_1$,  
\bea
&&
 (-ip_{0;l})[\prod_{v\in \cV} \delta(\sum_{l=1}^4p_{0;l;v})]  = 
\Big[-i\sum_{c \in {\rm Cycle}_\cG}\varepsilon_{l c}p_{0;c}
-i \sum_{  l_\ext } \varepsilon_{l l_\ext}p_{0;l_\ext}  \Big]
  [\prod_{v\in \cV} \delta(\sum_{l=1}^4p_{0;l;v})]  \cr\cr
&&
= 
 [-i\sum_{c \in {\rm Cycle}_\cG}\varepsilon_{l c}p_{0;c}]
 \Big[1 + Q^{4;1}_l  \Big]  [\prod_{v\in \cV} \delta(\sum_{l=1}^4p_{0;l;v})]  \,,\cr\cr
 &&
 Q^{4;1}_l= \frac{   \sum_{  l_\ext } \varepsilon_{l l_\ext}p_{0;l_\ext}
 }{\sum_{c \in {\rm Cycle}_\cG}\varepsilon_{l c}p_{0;c}}\,. 
 \label{ip01}
\eea
Then for elements  $l\in \cL_{\inter}\cap \cL_2$,  we write
\bea
&&
 e( \sum_{f\in \cF}\epsilon_{l f} p_f)[\prod_{v\in \cV} \delta(\sum_{l=1}^4p_{0;l;v})]  =
\Big[ e( \sum_{f\in \cF_\inter}\epsilon_{l f} p_f+ \sum_{f\in \cF_\ext}\epsilon_{l f} p^{\ext}_f) \Big]
  [\prod_{v\in \cV} \delta(\sum_{l=1}^4p_{0;l;v})]  \cr\cr
&&
= e( \sum_{f\in \cF_\inter}\epsilon_{l f} p_f)
 \Big[1 + Q^{4;2}_l  \Big]  [\prod_{v\in \cV} \delta(\sum_{l=1}^4p_{0;l;v})] \,, \cr\cr
 &&
 Q^{4;2}_l= \frac{   \sum_{f\in \cF_\ext}\epsilon_{l f} (p^{\ext}_f)^2  }{e( \sum_{f\in \cF_\inter}\epsilon_{l f} p_f)}\,. 
 \label{ip02}
\eea

We are in position to provide the local expansion of \eqref{ampl41}.  Plugging
\eqref{expl0}, \eqref{expf}, \eqref{ip01}  and \eqref{ip02}, in the four-point amplitude  \eqref{ampl41} we find: 
\bea\label{ampl42}
&&
A_{\cG;4}(\{p_{0;l_\ext}\};\{p^{\ext}_f\}) =\kappa(\lambda) \delta(\sum_{l_\ext} p_{0;l_\ext} )
 \int [\prod_{\ell\in \cL}\md\alpha_{\ell}  e^{\alpha_\ell}]
\int [\prod_{c \in {\rm Cycle}_\cG} \md p_{0;c}] [ \prod_{f \in \cF_{\inter}}\md p_{f}]\cr\cr
&&
\times 
\Big[\prod_{l \in \cL_\inter\cap \cL_1} [i\sum_{c \in {\rm Cycle}_\cG}\varepsilon_{l c}p_{0;c}]
[1 + Q^{4;1}_l]\Big]\Big[
\prod_{l \in \cL_\inter\cap \cL_2}e( \sum_{f\in \cF_\inter}\epsilon_{l f} p_f)[1 + Q^{4;2}_l]\Big] \cr\cr
&&
\Big[[\prod_{l_\ext \in \cL_\ext\cap \cL_1}  (-ip_{0;l_\ext})]
[\prod_{l_\ext \in \cL_\ext\cap \cL_2}  e( \sum_{f\in \cF_\ext}\epsilon_{l_\ext f} p^{\ext}_f)] \Big] \cr\cr
&&
\times 
[ \prod_{f \in \cF_{\ext} }e^{  -(\alpha_{ l_{\ext} } + \alpha_{ l_{\ext}' }) [(p^\ext_f)^{4} -2 (p^\ext_f)^{2}]   }
   (1- Q^1_{\ext ;f})] 
[ \prod_{\substack{f,f' \in \cF_{\ext}\\ f \ne f' }}e^{-(\alpha_{l_\ext}+\alpha_{l_\ext'}) (p^\ext_f)^{2} (p^\ext_{f'})^{2} }    (1- Q^2_{\ext ;f,f'}) ]\cr\cr
&& \times
\Big[ \prod_{f \in \cF_{\ext}, f' \in \cF_{\inter} }[ 1-Q^3_{\ext;f,f'}] \Big]
[  \prod_{ f \in \cF_{\inter} } e^{-(\sum_{\ell\in f} \alpha_{\ell})[ p^4_f - 2p^2_f ] }]
[ \prod_{\substack{f,f' \in \cF_{\inter} \\ f \ne f'}}e^{-(\sum_{\ell \in f, \ell \in f'}\alpha_\ell) (p_f)^{2} (p_{f'})^{2} } ]
\cr\cr
&&\times 
[\prod_{ c \in {\rm Cycle}_{\cG} }
 e^{- (\sum_{l \in \cL_c } \alpha_l)  p_{0;c}^2 } ] [ \prod_{\substack{c,c' \in {\rm Cycle}_{\cG} \\ c\ne c'}}
 e^{- (\sum_{ l \in \cT_c\cap \cT_{c'}  } \al_{l})p_{0;c}p_{0;c'} } ]
\crcr
 &&
 \times 
\Big[\prod_{l_\ext\in \cL_{\ext} }e^{- \alpha_{l_\ext} p_{0;l_{\ext}}^2  } [  1- Q^{1}_{l_\ext}] \Big]
\Big[ \prod_{l_\ext\in \cL_{\ext} }[  1- Q^{2}_{l_\ext}]\Big]
\Big[ \prod_{\substack{l_\ext,l_{\ext}'\in \cL_{\ext}\\ l_\ext\ne l_{\ext}'} }
[1 -  Q^{3}_{l_\ext,l_\ext'}]\Big]
\eea 
and this recasts as
\bea\label{ampli43}
&&
A_{\cG;4}(\{p_{0;l_\ext}\};\{p^{\ext}_f\}) =
\kappa(\lambda) \delta(\sum_{l_\ext} p_{0;l_\ext} )
 \int [\prod_{l_\ext\in \cL_{\ext}}\md\alpha_{l_\ext}  e^{\alpha_{l_\ext}}]\cr\cr
 &&\times 
 \Big[\prod_{l_\ext \in \cL_\ext \cap \cL_1}  (-ip_{0;l_\ext})  \Big]\Big[
\prod_{l_\ext \in \cL_\ext \cap \cL_2}  e( \sum_{f\in \cF_\ext}\epsilon_{l_\ext f} p^{\ext}_f) )\Big]
\cr\cr
&&\times
\Big[ \prod_{f \in \cF_{\ext} }e^{  -(\alpha_{ l_{\ext} } + \alpha_{ l_{\ext}' }) [(p^\ext_f)^{4} -2 (p^\ext_f)^{2}]   } \Big]\Big[ \prod_{\substack{f,f' \in \cF_{\ext}\\ f \ne f' }}e^{-(\alpha_{l_\ext}+\alpha_{l_\ext'}) (p^\ext_f)^{2} (p^\ext_{f'})^{2} }  \Big] 
\Big[\prod_{l_\ext\in \cL_{\ext} }e^{- \alpha_{l_\ext} p_{0;l_{\ext}}^2  } \Big]\cr\cr
&&\times
\int [\prod_{l\in \cL_{\inter}}\md\alpha_{l}  e^{\alpha_{l}}]
\int [\prod_{c \in {\rm Cycle}_\cG} \md p_{0;c}] [ \prod_{f \in \cF_{\inter}}\md p_{f}]
\Big[\prod_{l \in \cL_\inter\cap \cL_1} (-i\sum_{c \in {\rm Cycle}_\cG}\varepsilon_{l c}p_{0;c})\Big]\Big[\prod_{l \in \cL_\inter\cap \cL_2} e( \sum_{f\in \cF_\inter}\epsilon_{l f} p_f)\Big] 
\cr\cr
&& \times 
[  \prod_{ f \in \cF_{\inter} } e^{-(\sum_{\ell\in f} \alpha_{\ell})[ p^4_f - 2p^2_f ] }]
[ \prod_{\substack{f,f' \in \cF_{\inter} \\ f \ne f'}}e^{-(\sum_{\ell \in f, \ell \in f'}\alpha_\ell) (p_f)^{2} (p_{f'})^{2} } ]
\cr\cr
&&\times 
[\prod_{ c \in {\rm Cycle}_{\cG} }
 e^{- (\sum_{l \in \cL_c } \alpha_l)  p_{0;c}^2 } ] [ \prod_{\substack{c,c' \in {\rm Cycle}_{\cG} \\ c\ne c'}}
 e^{- (\sum_{ l \in \cT_c\cap \cT_{c'}  } \al_{l})p_{0;c}p_{0;c'} } ]
\crcr
&&
\times 
\Bigg\{ 1 +\sum_{\s=1,2} \sum_{l \in \cL_\inter\cap \cL_\s}Q^{4;\s}_l 
		- \sum_{f \in \cF_{\ext} } Q^1_{\ext ;f}
		- \sum_{\substack{f,f' \in \cF_{\ext}\\ f \ne f' }}Q^2_{\ext ;f,f'} 
		- \sum_{f \in \cF_{\ext}, f' \in \cF_{\inter} }Q^3_{\ext;f,f'} \cr\cr
&&
	- \sum_{l_\ext\in \cL_{\ext} } Q^{1}_{l_\ext} 
	- \sum_{l_\ext\in \cL_{\ext} }Q^{2}_{l_\ext}
 - \sum_{\substack{l_\ext,l_{\ext}'\in \cL_{\ext}\\ l_\ext\ne l_{\ext}'} }Q^{3}_{l_\ext,l_\ext'} 			
+ \sum Q \cdot Q + \dots \Bigg\} \,.
\eea
where  the last expression $ \sum Q \cdot Q + \dots$ stands for higher order
products of the remainders $Q$.  

The zeroth order in that expansion is of the form
\bea
&&
A_{\cG;4}(\{p_{0;l_\ext}\};\{p^{\ext}_f\};0) =
\kappa(\lambda) \delta(\sum_{l_\ext} p_{0;l_\ext} )
 \Big[\prod_{l_\ext \in \cL_\ext \cap \cL_1}  (-ip_{0;l_\ext})  \Big]\Big[
\prod_{l_\ext \in \cL_\ext \cap \cL_2}  e( \sum_{f\in \cF_\ext}\epsilon_{l_\ext f} p^{\ext}_f) )\Big]
  \label{prop1}  \\
 && \times \int [\prod_{l_\ext\in \cL_{\ext}}\md\alpha_{l_\ext}  e^{\alpha_{l_\ext}}]
 \Big[\prod_{l_\ext\in \cL_{\ext} }e^{- \alpha_{l_\ext} p_{0;l_{\ext}}^2  } \Big]
 \label{prop11}
\\
&&\times
\Big[ \prod_{f \in \cF_{\ext} }e^{  -(\alpha_{ l_{\ext} } + \alpha_{ l_{\ext}' }) [(p^\ext_f)^{4} -2 (p^\ext_f)^{2}]   } \Big]\Big[ \prod_{\substack{f,f' \in \cF_{\ext}\\ f \ne f' }}e^{-(\alpha_{l_\ext}+\alpha_{l_\ext'}) (p^\ext_f)^{2} (p^\ext_{f'})^{2} }  \Big] \label{prop2}\\
&&\times
\int [\prod_{l\in \cL_{\inter}}\md\alpha_{l}  e^{\alpha_{l}}]
\int [\prod_{c \in {\rm Cycle}_\cG} \md p_{0;c}] [ \prod_{f \in \cF_{\inter}}\md p_{f}]
\Big[\prod_{l \in \cL_\inter} [-i\sum_{c \in {\rm Cycle}_\cG}\varepsilon_{l c}p_{0;c}+ e( \sum_{f\in \cF_\inter}\epsilon_{l f} p_f)]\Big] 
\cr\cr
&& \times 
[  \prod_{ f \in \cF_{\inter} } e^{-(\sum_{\ell\in f} \alpha_{\ell})[ p^4_f - 2p^2_f ] }]
[ \prod_{\substack{f,f' \in \cF_{\inter} \\ f \ne f'}}e^{-(\sum_{\ell \in f, \ell \in f'}\alpha_\ell) (p_f)^{2} (p_{f'})^{2} } ]
\cr\cr
&&\times 
[\prod_{ c \in {\rm Cycle}_{\cG} }
 e^{- (\sum_{l \in \cL_c } \alpha_l)  p_{0;c}^2 } ] [ \prod_{\substack{c,c' \in {\rm Cycle}_{\cG} \\ c\ne c'}}
 e^{- (\sum_{ l \in \cT_c\cap \cT_{c'}  } \al_{l})p_{0;c}p_{0;c'} } ]\,.
\eea
By a small combinatorics and essentially variable renaming, the expressions \eqref{prop1}, \eqref{prop11} and \eqref{prop2} can be combined to give 4 propagators glued together to form a vertex with pattern given by 
\eqref{corr} and the three last lines  are integrals over internal momenta 
and will give a log-divergent contribution. This terms will therefore renormalize
$\lambda_m$ associated with the melonic vertex of the form $\bb_1$.

We now address the $Q$ remainder terms and recall that, for an internal line $l$
we have $p_{0;l} \sim M^{i_l} \sim \alpha_{l}^{-\frac12}$, for an external line
$l_\ext$, $p_{0;l_\ext} \sim M^{j_{l_\ext}} \sim \alpha_{l_\ext}^{-\frac12}$.
A momentum $p_{f}$ associated with a closed or external face $f$ is of the order $p_{f} \sim M^{-i_f/2}$, $i_f = \min_{\ell \in f } i_\ell $.  Note that if $f$ is external, 
then necessarily $i_f$ is nothing but one of the index $j_{l_\ext}$ of one of the two external sliced propagators $l_\ext$.

Keeping in mind $i(\cG^i_{(k)})= \min_{l\in \cL_{\inter}(\cG^i_{(k)})}i_l>e(\cG^i_{(k)})= \sup_{l_\ext \in \cL_{\ext}(\cG^i_{(k)})} j_{l_\ext}$, the following bounds 
are valid on a single $\cG_{(k)}^i$ graph: 
\bea
&& 
 | \sum_{l \in \cL_\inter \cap \cL_1}Q^{4;1}_l |
=   \sum_{l \in \cL_\inter}  
\frac{|\sum_{  l_\ext } \varepsilon_{l l_\ext}p_{0;l_\ext}| }{|\sum_{c \in {\rm Cycle}_\cG}\varepsilon_{l c}p_{0;c}|} 
 \leq   \frac{  c_1 M^{e(\cG^i_{(k)}) }  }{  c_2M^{ i(\cG^i_{(k)}) } } 
 \leq C_{4;1} M^{ -( i(\cG^i_{(k)}) - e(\cG^i_{(k)}) ) }   \,,  \cr\cr
  && 
 | \sum_{l \in \cL_\inter \cap \cL_2}Q^{4;2}_l |
=   \sum_{l \in \cL_\inter\cap \cL_2}  
\frac{ |\sum_{f\in \cF_\ext}\epsilon_{l f} (p^{\ext}_f)^2| }{| e( \sum_{f\in \cF_\inter}\epsilon_{l f} p_f) |}  
 \leq   \frac{  c_1' M^{e(\cG^i_{(k)}) } }{  c_2'M^{ i(\cG^i_{(k)}) } } 
  \leq C_{4;2} M^{ -( i(\cG^i_{(k)}) - e(\cG^i_{(k)}) ) }  \,,  \cr\cr
 &&
|\sum_{l_\ext \in \cL_\ext } Q^1_{l_\ext} |  = 
\Big| \sum_{l_\ext \in \cL_\ext } (\sum_{ l \in  \cT_{l_\ext} } \alpha_{l}) \, p_{0;l_{\ext}}^2 
 \int_0^1 \md s\,
 e^{- (\sum_{ l \in  \cT_{l_\ext} } \alpha_{l}) p_{0;l_{\ext}}^2  } \Big|   
 \leq  C_1 M^{-2(i(\cG^i_{(k)}) - e(\cG^i_{(k)}))}\,, \cr\cr
 &&
|\sum_{l_\ext \in \cL_\ext }  Q^{2}_{l_\ext}  |=2\Big|
\Big[\sum_{c \in  {\rm Cycle}_{\cG}}(\sum_{l \in \cT_c\cap \cT_{l_\ext}  } \al_{l}) p_{0;c}\Big] p_{0;l_{\ext}}
 \int_0^1 \md s\,
 e^{- 2 (\sum_{c \in  {\rm Cycle}_{\cG}}(\sum_{l \in \cT_c\cap \cT_{l_\ext}  } \al_{l}) p_{0;c}) p_{0;l_{\ext}} }\Big|  
 \cr\cr
 && \leq C_2 M^{-(2i(\cG^i_{(k)}) -i(\cG^i_{(k)})  - e(\cG^i_{(k)}))}  \leq C_2 M^{-( i(\cG^i_{(k)}) - e(\cG^i_{(k)}))} \,,
 \cr\cr
 && 
\Big|    \sum_{\substack{l_\ext,l_{\ext}'\in \cL_{\ext}\\ l_\ext\ne l_{\ext}'} }
  Q^{3}_{l_\ext, l'_{\ext}} \Big|  
   = \Big|  \sum_{\substack{l_\ext,l_{\ext}'\in \cL_{\ext}\\ l_\ext\ne l_{\ext}'} } (\sum_{ l \in \cT_{l_\ext}\cap \cT_{l'_\ext}  } \al_{l})p_{0;l_{\ext}}p_{0;l_{\ext}'}
    \int_0^1 \md s\,
    e^{- (\sum_{ l \in \cT_{l_\ext}\cap \cT_{l'_\ext}  } \al_{l})p_{0;l_{\ext}}p_{0;l_{\ext}'} } \Big|  
\cr\cr
&&
\leq  C_3 M^{-2(i(\cG^i_{(k)}) -e(\cG^i_{(k)})) }\,, 
\cr\cr
&&
\Big|   \sum_{f \in \cF_\ext} Q^1_{\ext ;f} \Big|  =   \Big|    \sum_{f \in \cF_\ext}   (\sum_{l \in f } \alpha_{l})  [(p^\ext_f)^{4} -2 (p^\ext_f)^{2}] 
   \int_0^{1} \md s\, e^{-s(\sum_{l \in f}\alpha_l )(p_f^{\ext})^2 }  \Big|   \cr\cr
   && 
   \leq C'_1  M^{-2(i(\cG^i_{(k)}) -e(\cG^i_{(k)})) }\,, \cr\cr
   &&
\Big|   \sum_{\substack{f,f' \in \cF_{\ext}\\ f \ne f' }}   Q^2_{\ext;f,f'} \Big|  =  
\Big|   \sum_{\substack{f,f' \in \cF_{\ext}\\ f \ne f' }}   ( \sum_{l \in f, l\in f' } \alpha_{l})  (p^\ext_f)^{2}(p^\ext_{f'})^{2}
   \int_0^{1} \md s\,  e^{-s(\sum_{l \in f, l\in f'}\alpha_l )(p_f^{\ext})^2(p_{f'}^{\ext})^2 }\Big|  \cr\cr
   && \leq   C'_2  M^{-2(i(\cG^i_{(k)}) -e(\cG^i_{(k)})) }\,, \cr\cr
   &&
    \Big|   \sum_{\substack{f,f' \in \cF_{\ext}\\ f \ne f' }}  
   Q^3_{\ext;f,f'}  \Big| =    \Big| ( \sum_{l \in f, l\in f' } \alpha_{l})  (p^\ext_f)^{2}(p_{f'})^{2}
   \int_0^{1}\md s\, e^{-s(\sum_{l \in f, l\in f'}\alpha_l )(p_f^{\ext})^2(p_{f'})^2 }  \Big| \cr\cr
   &&
     \leq C'_3  M^{-(2i(\cG^i_{(k)}) +i(\cG^i_{(k)}) - e(\cG^i_{(k)})) } \leq 
      C'_3  M^{-(i(\cG^i_{(k)})  - e(\cG^i_{(k)})) } \,,
 \eea
 where  $C_i$, $c_i$, $c_i'$ and $C'_i$ are constants depending on the graph. 
 Using these bounds, we have the following bound on the first order
corrections: 
\bea
|A_{\cG;4}(\{p_{0;l_\ext}\};\{p^{\ext}_f\};1) | \leq 
C  \prod_{(i,k)\in \mathbb{N}^2} M^{\omega_{\md}(\cG^i_{(k)})}
M^{-  (i(\cG^i_{(k)})  - e(\cG^i_{(k)})}\,,
\eea
where $C$ is another constant. Hence, since $i(\cG^i_{(k)})  - e(\cG^i_{(k)})>0$, 
this bound shows that the remainder will bring enough decay to 
ensure the convergence during the sum over scale attributions. 
In the same vein, higher order products of $Q^{(\cdot)}$'s will be 
even more convergent. Finally, after changing the pattern of external momenta in the four-point 
correlator in a way to produce other type of melonic interactions of the form $\bb_c$ of any color $c=0,1,2,3$, 
we can perform an analysis entirely parallel to the above and show that the 
zeroth order term will renormalize $\lambda_m$ and remainders will be again 
convergent.

\medskip
\noindent{\bf Two-point amplitudes.} 
There is a unique boundary graph for any two-point amplitude and it is such that  $g_{\bG} =0$. 
As discussed in section \ref{sect:renorm}, there are several types of two-point graphs 
which could diverge. Their general degree of divergence is of the form $\omega_{\deg}(\cG)=1-p/2$, $p\in \{0,1,2\}$. 
We will focus on the maximal degree case, that is $p=0, \omega_{\deg}(\cG)=1$, where the expansion needs to be
pushed at second order. There other cases can be understood from this point.  

We consider a perturbative amplitude issued from the expansion if the correlator  
\be\label{corr2}
\langle \psi_{p_{0;1}; 123} \,\psi_{p_{0;1}123} \rangle \,, 
\ee
where $\psi= \chi_\sigma$. 
The expression \eqref{ampl41}  remains true for any graph amplitude.  We now expand the exponentials appearing therein: 
\bea
&&
 e^{- (\alpha_{l_\ext} +\sum_{ l \in  \cT_{l_\ext} } \alpha_{l}) p_{0;l_{\ext}}^2  } 
  =
 e^{- \alpha_{l_\ext} p_{0;l_{\ext}}^2  } 
 \Big[  1- Q^{1}_{l_\ext} + Q^{1'}_{l_\ext}\Big] \,,\cr\cr
 &&
 Q^1_{l_\ext} = 
 (\sum_{ l \in  \cT_{l_\ext} } \alpha_{l}) \, p_{0;l_{\ext}}^2\,,  \qquad
 Q^{1'}_{l_\ext} = 
 [(\sum_{ l \in  \cT_{l_\ext} } \alpha_{l}) \, p_{0;l_{\ext}}^2 ]^2 \int_0^1 \md s\,(s-1)
 e^{-s (\sum_{ l \in  \cT_{l_\ext} } \alpha_{l}) p_{0;l_{\ext}}^2  }\,,  \cr\cr
 && 
 e^{- 2 (\sum_{c \in  {\rm Cycle}_{\cG}}(\sum_{l \in \cT_c\cap \cT_{l_\ext}  } \al_{l}) p_{0;c}) p_{0;l_{\ext}} } 
  = 1 -  Q^{2}_{l_\ext} + Q^{2'}_{l_\ext} \,,\cr\cr
 && 
Q^{2}_{l_\ext}   =2
\Big[\sum_{c \in  {\rm Cycle}_{\cG}}(\sum_{l \in \cT_c\cap \cT_{l_\ext}  } \al_{l}) p_{0;c}\Big] p_{0;l_{\ext}} \,,\cr\cr
&&
Q^{2'}_{l_\ext} = 
\Big[2\Big[\sum_{c \in  {\rm Cycle}_{\cG}}(\sum_{l \in \cT_c\cap \cT_{l_\ext}  } \al_{l}) p_{0;c}\Big] p_{0;l_{\ext}} \Big]^2
 \int_0^1 \md s\,(1-s)
 e^{- 2 s(\sum_{c \in  {\rm Cycle}_{\cG}}(\sum_{l \in \cT_c\cap \cT_{l_\ext}  } \al_{l}) p_{0;c}) p_{0;l_{\ext}} } \,, \cr\cr
 &&
 e^{- (\sum_{ l \in \cT_{l_\ext}\cap \cT_{l'_\ext}  } \al_{l})p_{0;l_{\ext}}p_{0;l_{\ext}'} } 
  =  1 -    Q^{3}_{l_\ext, l'_{\ext}}  +  Q^{3'}_{l_\ext, l'_{\ext}} \,, \cr\cr
  &&
  Q^{3}_{l_\ext, l'_{\ext}} 
   = (\sum_{ l \in \cT_{l_\ext}\cap \cT_{l'_\ext}  } \al_{l})p_{0;l_{\ext}}p_{0;l_{\ext}'}\,, \cr\cr
   && 
     Q^{3'}_{l_\ext, l'_{\ext}} =
     \Big[(\sum_{ l \in \cT_{l_\ext}\cap \cT_{l'_\ext}  } \al_{l})p_{0;l_{\ext}}p_{0;l_{\ext}'}\Big]^2 
    \int_0^1 \md s\,(1-s)
    e^{- s(\sum_{ l \in \cT_{l_\ext}\cap \cT_{l'_\ext}  } \al_{l})p_{0;l_{\ext}}p_{0;l_{\ext}'} } \,. 
    \label{expl02}
\eea
Meanwhile, for momenta associated with faces, we have 
\bea
&& 
e^{-(\sum_{\ell \in f}\alpha_\ell) [(p^\ext_f)^{4} -2 (p^\ext_f)^{2}] }
  =  e^{  -(\alpha_{ l_{\ext} } + \alpha_{ l_{\ext}' }) [(p^\ext_f)^{4} -2 (p^\ext_f)^{2}]   }
   (1- Q^1_{\ext ;f}+  Q^{1'}_{\ext ;f})\,,
 \cr\cr
 &&  
 Q^1_{\ext ;f} =     (\sum_{l \in f } \alpha_{l})  [(p^\ext_f)^{4} -2 (p^\ext_f)^{2}]\,, \cr\cr
 && 
 Q^{1'}_{\ext ;f} =    \Big[ (\sum_{l \in f } \alpha_{l})  [(p^\ext_f)^{4} -2 (p^\ext_f)^{2}] \Big]^2 
   \int_0^{1} \md s\, (1-s) e^{-s(\sum_{l \in f}\alpha_l )(p_f^{\ext})^2 }  \,,    \cr\cr
   && 
 e^{-(\sum_{\ell \in f, \ell \in f'}\alpha_\ell) (p^\ext_f)^{2} (p^\ext_{f'})^{2} } 
 =  e^{-(\alpha_{l_\ext}+\alpha_{l_\ext'}) (p^\ext_f)^{2} (p^\ext_{f'})^{2} }    (1- Q^2_{\ext ;f}+ Q^{2'}_{\ext ;f})\,,
 \cr\cr
 && 
 Q^2_{\ext;f,f'} =     ( \sum_{l \in f, l\in f' } \alpha_{l})  (p^\ext_f)^{2}(p^\ext_{f'})^{2}\,,\cr\cr
 &&
 Q^{2'}_{\ext;f,f'} =  \Big[ ( \sum_{l \in f, l\in f' } \alpha_{l})  (p^\ext_f)^{2}(p^\ext_{f'})^{2} \Big]^2
   \int_0^{1} \md s\, (1-s) e^{-s(\sum_{l \in f, l\in f'}\alpha_l )(p_f^{\ext})^2(p_{f'}^{\ext})^2 }  \,, 
   \cr\cr
   && 
  e^{-(\sum_{\ell \in f, \ell \in f'}\alpha_\ell) (p^\ext_f)^{2} (p_{f'})^{2} } 
  = 1-Q^3_{\ext;f,f'} +Q^{3'}_{\ext;f,f'} \,,  \cr\cr
  &&
  Q^3_{\ext;f,f'} =   ( \sum_{l \in f, l\in f' } \alpha_{l})  (p^\ext_f)^{2}(p_{f'})^{2} \,,\cr\cr
  &&
  Q^3_{\ext;f,f'} = \Big[( \sum_{l \in f, l\in f' } \alpha_{l})  (p^\ext_f)^{2}(p_{f'})^{2} \Big]^2
   \int_0^{1}\md s\, (1-s)e^{-s(\sum_{l \in f, l\in f'}\alpha_l )(p_f^{\ext})^2(p_{f'})^2 }\,.
       \label{expfff}
  \eea
The last factor to expand becomes: 
\bea
&&
[\prod_{l \in \cL_{\inter}\cap \cL_1} (-ip_{0;l})]
[\prod_{l \in \cL_{\inter}\cap \cL_2}
  e( \sum_{f\in \cF}\epsilon_{l f} p_f)] [\prod_{v\in \cV} \delta(\sum_{l=1}^4p_{0;l;v})]  = 
 [Q^1 + Q^{2;1}+Q^{2;2} + Q^3 ] [\prod_{v\in \cV} \delta(\sum_{l=1}^4p_{0;l;v})]\,, \cr\cr
 &&
Q^1 = [\prod_{l \in \cL_{\inter}\cap \cL_1}
(-i\sum_{c \in {\rm Cycle}_\cG}\varepsilon_{l c}p_{0;c})]
 [\prod_{l \in \cL_{\inter}\cap \cL_2} e( \sum_{f\in \cF_\inter}\epsilon_{l f} p_f) ] \,,\cr\cr
&& 
Q^{2;1}=  
\sum_{l \in \cL_{\inter}\cap \cL_1}\Big\{ \Big[-i \sum_{  l_\ext } \varepsilon_{l l_\ext}p_{0;l_\ext}  
\Big] \Big[
\prod_{\substack{l' \in \cL_{\inter}\cap \cL_1 \\ l'\neq l}}(
-i\sum_{c \in {\rm Cycle}_\cG}\varepsilon_{l' c}p_{0;c})\Big] \Big\} \Big[
\prod_{l \in \cL_{\inter}\cap \cL_2} e( \sum_{f\in \cF_\inter}\epsilon_{l' f} p_f)\Big] \,,\cr\cr
&&
Q^{2;2}=  
\sum_{l \in \cL_{\inter}\cap \cL_2}\Big\{ \Big[
\sum_{f\in \cF_\ext}\epsilon_{l f} (p^{\ext}_f)^2
\Big] \Big[
\prod_{\substack{l' \in \cL_{\inter}\cap \cL_2\\ l'\neq l}}
e( \sum_{f\in \cF_\inter}\epsilon_{l' f} p_f)\Big] \Big\}\Big[
\prod_{ l' \in \cL_{\inter}\cap \cL_1 }(-i\sum_{c \in {\rm Cycle}_\cG}\varepsilon_{l' c}p_{0;c})\Big]\,,  
 \label{ipoll}
\eea  
 and $Q^3$ is the sum of all remainder terms invoking all higher orders of the product 
 $$\Big|\prod_{l \in A}\Big[-i \sum_{  l_\ext } \varepsilon_{l l_\ext}p_{0;l_\ext}  
\Big]\prod_{l \in B}\Big[ \sum_{f\in \cF_\ext}\epsilon_{l f} (p^{\ext}_f)^2\Big]\Big|,$$ 
for two subsets $A$ and $B$ of internal 
lines, $A,B\subset \cL_{\inter}$,  with cardinality $|A| \geq 2$ if $|B|=0$, or $|B|>1$ if $|A| =0$, 
or $|A|+2|B| \ge 3$, if $A>0$ and $B>0$. 

We insert these expansions in the 2-point amplitude and get: 
\bea\label{ampli2}
&&
A_{\cG;2}(\{p_{0;l_\ext}\};\{p^{\ext}_f\}) =
\kappa(\lambda) \delta(\sum_{l_\ext} p_{0;l_\ext} )
 \int [\prod_{l_\ext\in \cL_{\ext}}\md\alpha_{l_\ext}  e^{\alpha_{l_\ext}}]\cr\cr
 &&\times 
 \Big[\prod_{l_\ext \in \cL_\ext \cap \cL_1}  (-ip_{0;l_\ext})  \Big]\Big[
\prod_{l_\ext \in \cL_\ext \cap \cL_2}  e( \sum_{f\in \cF_\ext}\epsilon_{l_\ext f} p^{\ext}_f) \Big]
\cr\cr
&&\times
\Big[ \prod_{f \in \cF_{\ext} }e^{  -(\alpha_{ l_{\ext} } + \alpha_{ l_{\ext}' }) [(p^\ext_f)^{4} -2 (p^\ext_f)^{2}]   } \Big]\Big[ \prod_{\substack{f,f' \in \cF_{\ext}\\ f \ne f' }}e^{-(\alpha_{l_\ext}+\alpha_{l_\ext'}) (p^\ext_f)^{2} (p^\ext_{f'})^{2} }  \Big] 
\Big[\prod_{l_\ext\in \cL_{\ext} }e^{- \alpha_{l_\ext} p_{0;l_{\ext}}^2  } \Big]\cr\cr
&& \times 
[  \prod_{ f \in \cF_{\inter} } e^{-(\sum_{\ell\in f} \alpha_{\ell})[ p^4_f - 2p^2_f ] }]
[ \prod_{\substack{f,f' \in \cF_{\inter} \\ f \ne f'}}e^{-(\sum_{\ell \in f, \ell \in f'}\alpha_\ell) (p_f)^{2} (p_{f'})^{2} } ]
\cr\cr
&&\times 
[\prod_{ c \in {\rm Cycle}_{\cG} }
 e^{- (\sum_{l \in \cL_c } \alpha_l)  p_{0;c}^2 } ] [ \prod_{\substack{c,c' \in {\rm Cycle}_{\cG} \\ c\ne c'}}
 e^{- (\sum_{ l \in \cT_c\cap \cT_{c'}  } \al_{l})p_{0;c}p_{0;c'} } ]\Big(Q^{1}+\sum_{\s=1,2}Q^{2;\s}+Q^{3}\Big)
\crcr
&&
\times 
\Bigg\{ 1 
		- \sum_{f \in \cF_{\ext} }( Q^1_{\ext ;f} + Q^{1'}_{\ext ;f})
		- \sum_{\substack{f,f' \in \cF_{\ext}\\ f \ne f' }}(Q^2_{\ext ;f,f'} +Q^{2'}_{\ext ;f,f'} )
		- \sum_{f \in \cF_{\ext}, f' \in \cF_{\inter} }(Q^3_{\ext ;f,f'} +Q^{3'}_{\ext ;f,f'} ) \cr\cr
&&
	- \sum_{l_\ext\in \cL_{\ext} } (Q^{1}_{l_\ext} +Q^{1'}_{l_\ext})
	- \sum_{l_\ext\in \cL_{\ext} }(Q^{2}_{l_\ext} +Q^{2'}_{l_\ext})
 - \sum_{\substack{l_\ext,l_{\ext}'\in \cL_{\ext}\\ l_\ext\ne l_{\ext}'} }(Q^{3}_{l_\ext,l_\ext'} 	+Q^{3'}_{l_\ext,l_\ext'} )	\cr\cr
 && 	 
+ \sum (Q+Q) \cdot( Q +Q) + \dots \Bigg\} \,.
\eea
where   $ \sum (Q+Q) \cdot( Q +Q) + \dots$ involves all types of higher order
products of the remainders $Q^{(\cdot)}_{\ext;-}$ and 
$Q^{(\cdot)}_{l_\ext}$, $Q^{(\cdot)}_{l_\ext,l'_\ext}$ .  

At zeroth order, we have the following amplitude
\bea\label{ampli20}
&&
A_{\cG;2}(\{p_{0;l_\ext}\};\{p^{\ext}_f\};0) =
\kappa(\lambda) \delta(\sum_{l_\ext} p_{0;l_\ext} )
 \int [\prod_{l_\ext\in \cL_{\ext}}\md\alpha_{l_\ext}  e^{\alpha_{l_\ext}}] \cr\cr
 &&
\times  \Big[\prod_{l_\ext \in \cL_\ext \cap \cL_1}  (-ip_{0;l_\ext})  \Big]\Big[
\prod_{l_\ext \in \cL_\ext \cap \cL_2}  e( \sum_{f\in \cF_\ext}\epsilon_{l_\ext f} p^{\ext}_f )\Big]
\cr\cr
&&\times
\Big[ \prod_{f \in \cF_{\ext} }e^{  -(\alpha_{ l_{\ext} } + \alpha_{ l_{\ext}' }) [(p^\ext_f)^{4} -2 (p^\ext_f)^{2}]   } \Big]\Big[ \prod_{\substack{f,f' \in \cF_{\ext}\\ f \ne f' }}e^{-(\alpha_{l_\ext}+\alpha_{l_\ext'}) (p^\ext_f)^{2} (p^\ext_{f'})^{2} }  \Big] 
\Big[\prod_{l_\ext\in \cL_{\ext} }e^{- \alpha_{l_\ext} p_{0;l_{\ext}}^2  } \Big]\cr\cr
&&\times
\int [\prod_{l\in \cL_{\inter}}\md\alpha_{l}  e^{\alpha_{l}}]
\int [\prod_{c \in {\rm Cycle}_\cG} \md p_{0;c}] [ \prod_{f \in \cF_{\inter}}\md p_{f}]
 [\prod_{l \in \cL_{\inter}\cap \cL_1}
(-i\sum_{c \in {\rm Cycle}_\cG}\varepsilon_{l c}p_{0;c})]
 [\prod_{l \in \cL_{\inter}\cap \cL_2} e( \sum_{f\in \cF_\inter}\epsilon_{l f} p_f) ]
\cr\cr
&& \times 
[  \prod_{ f \in \cF_{\inter} } e^{-(\sum_{\ell\in f} \alpha_{\ell})[ p^4_f - 2p^2_f ] }]
[ \prod_{\substack{f,f' \in \cF_{\inter} \\ f \ne f'}}e^{-(\sum_{\ell \in f, \ell \in f'}\alpha_\ell) (p_f)^{2} (p_{f'})^{2} } ]
\cr\cr
&&\times 
[\prod_{ c \in {\rm Cycle}_{\cG} }
 e^{- (\sum_{l \in \cL_c } \alpha_l)  p_{0;c}^2 } ] [ \prod_{\substack{c,c' \in {\rm Cycle}_{\cG} \\ c\ne c'}}
 e^{- (\sum_{ l \in \cT_c\cap \cT_{c'}  } \al_{l})p_{0;c}p_{0;c'} } ] \,.
\eea
Some change of variables allows us  to show that the
contribution of the  external momenta can be recast as two propagators glued
together and the factors from the integral over internal momenta which produces a linearly 
divergent term. This term renormalizes the mass (or the chemical 
potential, hence the Fermi radius in a condensed matter interpretation).  Beware
that this mass renormalization has a logarithmically divergent part corresponding to the constant part of the $Q^1$ term in \eqref{ampli2}. 

We focus on the next order that we denote: 
\bea\label{ampli21}
&&
A_{\cG;2}(\{p_{0;l_\ext}\};\{p^{\ext}_f\};1) =
\kappa(\lambda) \delta(\sum_{l_\ext} p_{0;l_\ext} )
 \int [\prod_{l_\ext\in \cL_{\ext}}\md\alpha_{l_\ext}  e^{\alpha_{l_\ext}}] \cr\cr
&&
\times  \Big[\prod_{l_\ext \in \cL_\ext \cap \cL_1}  (-ip_{0;l_\ext})  \Big]\Big[
\prod_{l_\ext \in \cL_\ext \cap \cL_2}  e( \sum_{f\in \cF_\ext}\epsilon_{l_\ext f} p^{\ext}_f)\Big]
\cr\cr
&&\times
\Big[ \prod_{f \in \cF_{\ext} }e^{  -(\alpha_{ l_{\ext} } + \alpha_{ l_{\ext}' }) [(p^\ext_f)^{4} -2 (p^\ext_f)^{2}]   } \Big]\Big[ \prod_{\substack{f,f' \in \cF_{\ext}\\ f \ne f' }}e^{-(\alpha_{l_\ext}+\alpha_{l_\ext'}) (p^\ext_f)^{2} (p^\ext_{f'})^{2} }  \Big] 
\Big[\prod_{l_\ext\in \cL_{\ext} }e^{- \alpha_{l_\ext} p_{0;l_{\ext}}^2  } \Big]\cr\cr
&&\times 
\int [\prod_{l\in \cL_{\inter}}\md\alpha_{l}  e^{\alpha_{l}}]
\int [\prod_{c \in {\rm Cycle}_\cG} \md p_{0;c}] [ \prod_{f \in \cF_{\inter}}\md p_{f}]
[  \prod_{ f \in \cF_{\inter} } e^{-(\sum_{\ell\in f} \alpha_{\ell})[ p^4_f - 2p^2_f ] }]
[ \prod_{\substack{f,f' \in \cF_{\inter} \\ f \ne f'}}e^{-(\sum_{\ell \in f, \ell \in f'}\alpha_\ell) (p_f)^{2} (p_{f'})^{2} } ]
\cr\cr
&&\times 
[\prod_{ c \in {\rm Cycle}_{\cG} }
 e^{- (\sum_{l \in \cL_c } \alpha_l)  p_{0;c}^2 } ] [ \prod_{\substack{c,c' \in {\rm Cycle}_{\cG} \\ c\ne c'}}
 e^{- (\sum_{ l \in \cT_c\cap \cT_{c'}  } \al_{l})p_{0;c}p_{0;c'} } ]
\crcr
&&
\times 
\Bigg\{  \sum_{\s=1,2} Q^{2;\s} + Q^{1}\Big[ - \sum_{f \in \cF_{\ext}, f' \in \cF_{\inter} } Q^3_{\ext ;f,f'}   
	- \sum_{l_\ext\in \cL_{\ext} } Q^{2}_{l_\ext}  \Big]  \Bigg\} \,.
\eea
We focus on the $Q$'s terms and put them in the form
\bea
&&
Q^{1} \sum_{f \in \cF_{\ext}, f' \in \cF_{\inter} } Q^3_{\ext ;f,f'} 
     = \sum_{f \in \cF_{\ext} }  (p^\ext_f)^{2} \Big[\Big(\sum_{ f' \in \cF_{\inter} } (p_{f'})^{2}\Big) Q^1   ( \sum_{l \in f, l\in f' } \alpha_{l}) \Big] \cr\cr
&&
Q^{1}  \sum_{l_\ext\in \cL_{\ext} } Q^{2}_{l_\ext}  =
 \sum_{l_\ext\in \cL_{\ext} } p_{0;l_{\ext}}\Big[ 2Q^{1}
 \sum_{c \in  {\rm Cycle}_{\cG}}(\sum_{l \in \cT_c\cap \cT_{l_\ext}  } \al_{l}) p_{0;c}\Big] \cr\cr
 &&
 Q^{2;1}=  -i \sum_{  l_\ext \in \cL_{\ext} }p_{0;l_\ext}\Big\{ \sum_{l \in \cL_{\inter}\cap \cL_1}\varepsilon_{l l_\ext}
\prod_{\substack{l' \in \cL_{\inter}\cap \cL_1 \\ l'\neq l}}\Big[
-i\sum_{c \in {\rm Cycle}_\cG}\varepsilon_{l' c}p_{0;c}\Big] \Big\} \Big[
\prod_{l \in \cL_{\inter}\cap \cL_2} e( \sum_{f\in \cF_\inter}\epsilon_{l' f} p_f)\Big]\cr\cr
&&
 Q^{2;2}=
 \sum_{f\in \cF_\ext} (p^{\ext}_f)^2 \Big\{ 
 \sum_{l \in \cL_{\inter}\cap \cL_2} \epsilon_{l f}
\prod_{\substack{l' \in \cL_{\inter}\cap \cL_2\\ l'\neq l}}\Big[
e( \sum_{f\in \cF_\inter}\epsilon_{l' f} p_f)\Big] \Big\}
\Big[
\prod_{ l' \in \cL_{\inter}\cap \cL_1 }(-i\sum_{c \in {\rm Cycle}_\cG}\varepsilon_{l' c}p_{0;c})\Big]   \,. 
\eea
At this point, one observes that the integral over all internal momenta of the above expressions could be brought
as $ \sum_{f \in \cF_{\ext} }  (p^\ext_f)^{2} \times {\rm coeff}(f)$ and $-i p_{0;l_\ext} \times {\rm coeff}'(l_\ext)$, 
where ${\rm coeff}(f)$ and ${\rm coeff}'(l_\ext)$ are constants depending on the graph. 
To be able to put those results as 
\be
-i p_{0;l_\ext} {\rm coeff}'(l_\ext) + [\sum_{f \in \cF_{\ext} }  (p^\ext_f)^{2} ]\times {\rm coeff}
 = -i p_{0;l_\ext} {\rm coeff}'(l_\ext) +   [(p_{1;l_\ext})^2+(p_{2;l_\ext})^2+(p_{3;l_\ext})^{2} ]\times {\rm coeff},
\ee
which is of the form of the prefactor of the kinetic term and where coeff is another constant independent of $f$, 
we must gather all colored graphs which only differ through color permutation,
and sum their contributions which must be all equal.  
Thus this term (and the like by symmetrizing the graph) renormalize the two 
wave-functions
$\Delta_{p_0}$ and $\Delta_{p^2}$. 

The last step is to prove the convergence of all remainder terms. 
We provide the following bounds  of the remainders $Q$ (under bounded integrals)
\bea
&& 
\Big|   \sum_{f \in \cF_{\ext} }Q^1_{\ext ;f} \Big| 
   =   \Big| \sum_{f \in \cF_{\ext} } [(p^\ext_f)^{4} -2 (p^\ext_f)^{2}] 
    \Big[    (\sum_{l \in f } \alpha_{l})
\Big]\Big|  
  \leq k_1 M^{-2[i (\cG^i_{(k)}) - e (\cG^i_{(k)})]} \,,\cr\cr
&&
\Big|   \sum_{\substack{f,f' \in \cF_{\ext}\\ f \ne f' }}Q^2_{\ext ;f,f'}\Big|  = 
\Big|  \sum_{\substack{f,f' \in \cF_{\ext}\\ f \ne f' }}
   (p^\ext_f)^{2}(p^\ext_{f'})^{2} \Big[      ( \sum_{l \in f, l\in f' } \alpha_{l}) \Big]\Big|   \leq k_2 M^{-2[i (\cG^i_{(k)}) - e (\cG^i_{(k)})]}  \, ,\cr\cr
&&
\Big|   \sum_{l_\ext\in \cL_{\ext} } Q^{1}_{l_\ext} \Big| = 
\Big| \sum_{l_\ext\in \cL_{\ext} } (p_{0;l_{\ext}})^2\Big[   (\sum_{ l \in  \cT_{l_\ext} } \alpha_{l}) \Big] \Big|  
\leq k_3 M^{-2[i (\cG^i_{(k)}) - e (\cG^i_{(k)})]}\,, \cr\cr
&&
\Big|    \sum_{\substack{l_\ext,l_{\ext}'\in \cL_{\ext}\\ l_\ext\ne l_{\ext}'} } Q^{3}_{l_\ext,l_\ext'} \Big|	 = 
\Big| \sum_{\substack{l_\ext,l_{\ext}'\in \cL_{\ext}\\ l_\ext\ne l_{\ext}'} }
p_{0;l_{\ext}}p_{0;l_{\ext}'} 
\Big[ (\sum_{ l \in \cT_{l_\ext}\cap \cT_{l'_\ext}  } \al_{l})\Big] \Big| \leq 
k_4 M^{-2[i (\cG^i_{(k)}) - e (\cG^i_{(k)})]}\,,
 \eea
 where $k$'s are constants.  Using $i (\cG^i_{(k)}) - e (\cG^i_{(k)})>1$, and the fact that the integral over $Q^1$ brings the mass divergence $M^{ \omega_{\deg} (\cG^i_{(k)}) =1}$
 and that the integral over $Q^2$ 
brings lead to a log-divergent contribution, these bounds shows that any term 
in the expansion involving one of the above expression as a factor has a strictly negative divergence degree. Further, we have 
\bea
&& 
| Q^{1'}_{l_\ext}| \le k_5 M^{-4[i (\cG^i_{(k)})-e (\cG^i_{(k)})]} \,,  \cr\cr
 && 
|Q^{2'}_{l_\ext} |\le k_6  M^{-2[i (\cG^i_{(k)})-  e (\cG^i_{(k)})]} \,, \cr\cr
&&
|  Q^{3'}_{l_\ext, l'_{\ext}} | \le  k_7  M^{-4 [ i (\cG^i_{(k)})-  e (\cG^i_{(k)}) ] }  \,,\cr\cr
 && 
| Q^{1'}_{\ext ;f} | \le  k_7  M^{-4 [ i (\cG^i_{(k)})-  e (\cG^i_{(k)}) ] }  \,,    \cr\cr
 &&
| Q^{2'}_{\ext;f,f'} | \le  k_8  M^{-4 [ i (\cG^i_{(k)})-  e (\cG^i_{(k)}) ] }   \,, \cr\cr
  &&
 |  Q^3_{\ext;f,f'} | \le k_9 M^{2[  -2 i (\cG^i_{(k)})+ i (\cG^i_{(k)}) + e (\cG^i_{(k)})   ]} \le M^{-2[   i (\cG^i_{(k)})- e (\cG^i_{(k)})   ]}\,.
  \eea
Hence, any product of the above with $Q^1$ or $Q^{2;\s}$ will immediately lead  a negative degree of divergence. 
In the same way, we can also show that $Q^3$ and the other  higher order products of $Q$'s
will contribute to convergent terms. After removing the divergences, all these contributions bring
a sufficient decay to sum over the scale attributions and will lead to convergence. 
Thus,  the model becomes renormalizable at all orders of perturbations.

\section{Conclusion}
\label{conclusion}

We have proved the renormalizability of a tensor SYK model 
with a pair of Majorana tensor fields, in which time and tensor indices both govern a kind of renormalization group 
 $(t_0,\vec x)\in [- \frac{\beta}{2}, \frac{\beta}{2}] \times \R^3$ or $\in
 [- \frac{\beta}{2}, \frac{\beta}{2}] \times U(1)^3$. Our model considers the orthogonal invariant (melonic and tetraedric) interactions  introduced
by Carrozza-Tanasa and uses the local-time interaction 
introduced by Klebanov-Tarnolposki which is common to all the SYK-type models. 
But it is endowed with a new notion of renormalization since it is based on the standard propagator of non relativistic 
condensed matter. We achieved the proof of the perturbative ultra-violet renormalizability of the model through a multi-scale analysis
and a power counting theorem which, interestingly, mixes
the ordinary power counting of local field theory and the power counting of a non-local 
part coming from the tensorial convolution of the indices. 
A detailed study of the degree of divergence of 
an arbitrary graph proves that only the quartic melonic interactions
renormalize as expected from the large $N$ limit. 

Having shown perturbative ultra-violet renormalizability, a next step is to compute
the perturbative and non-perturbative flow equations for this model. 
Quartic melonic tensor field theory are generally UV asymptotically free \cite{BenGeloun:2012pu}. A
natural question is to check if this remains true for the tensor SYK field theories introduced in this paper. 
Here the model is somehow different with two wave function couplings ($\Delta_{p_0}$
and $\Delta_{p^2}$). The property of asymptotic safety or asymptotic freedom in the UV for tensor field theories 
mainly rests on the existence of a rapid growth of the coefficients of the wave function 
renormalization relatively to the quartic coupling. For our present situation, we foresee that, at one-loop,  the tadpole gives no contribution to $\Delta_{p_0}$ but there will be still a contribution to $\Delta_{p^2}$. All ingredients which trigger asymptotic freedom are therefore still present.

Of course the most interesting physics of this model lies in the infrared regime, 
which we intend to explore in a future study.
We expect the tetraedric interaction to become more interesting in this regime. 
We may also have to to consider variants 
of the model action \eqref{act:tens}, obtained  by coloring differently 
the vertices with the two fields $\chi_1$ and  $\chi_2$. 

\

\noindent{\bf Acknowledgments}
V. Rivasseau thanks N. Delporte, F. Ferrari, R. Gurau and G. Valette for useful discussions.

\end{document}